\documentclass[a4paper,11pt]{article}
\usepackage[latin1]{inputenc}
\usepackage{indentfirst}
\usepackage[left]{lineno}
\usepackage{amsfonts}
\usepackage{float}

\usepackage{latexsym,amssymb,amsmath}
\usepackage[dvips]{graphics,graphicx,epsfig,color}
\usepackage[lofdepth,lotdepth]{subfig}
\usepackage[lofdepth,lotdepth]{subfig}
\begin{document}
\title{\bf A conservation law for testing methods of prediction of the seismic wave response of a protuberance emerging from flat ground}
\author{Armand Wirgin\thanks{LMA, CNRS, UPR 7051, Aix-Marseille Univ, Centrale Marseille, F-13453 Marseille Cedex 13, France, ({\tt wirgin@lma.cnrs-mrs.fr})} }
\date{\today}
\maketitle
\begin{abstract}
We establish  the  equations which translate a conservation  law for the problem of the seismic response of an above-ground  structure (e.g., building, hill or mountain) of arbitrary shape and inquire whether both the implicit (formal) and explicit (numerical) solutions for the response obey this law for the case of a cylindrical, rectangular protuberance. Both the low-order (poor approximations of the response) as well as  higher-order (supposedly better approximations) turn out to satisfy the conservation of flux relation, which means that the satisfaction of this relation is a necessary, but not sufficient, means for determining whether a solution to the scattering problem is valid.
\end{abstract}
\newline
\newline
Abbreviated title: Conservation of flux relative to the seismic response of a protuberance
\newline
\newline
Corresponding author: Armand Wirgin, \\ e-mail: wirgin@lma.cnrs-mrs.fr
\newpage
\tableofcontents
\newpage
\newpage
\section{Introduction}\label{intro}
 Seismic waves  are known to damage  buildings  or industrial facilities located either on flat or hilly ground and thought to provoke land and rock slides on hills and mountains \cite{bf14,ck15,co14,dw73,dm10,gr09,hj02,he83,ho12,ke94,lh99,lo17,mn75,pa11,pr14,rr12,sm05,zs13}. These man-made or natural features can be grouped into the term: above-ground structure (AGS) or protuberance for short. The seismic response of protuberances is an ongoing research theme in theoretical and applied seismology, mainly because of the variety of AGS's, the social and economic impact of the damage issue and a relatively-poor understanding of the  empirical evidence that the seismic wave field is amplified on the stress-free boundary of the protuberance relative to this field on flat ground. For these reasons,  a great variety of (mainly-numerical) solutions to  specific AGS seismic response problems have been proposed \cite{as14,bm14,bb96,bf14,ck15,ca95,dm10,gr05,hj02,ho12,jc00,kw92,la13,lc10,lo17,ls95,mn75,pa11,ps94,ql05,rr12,rk08,rk74,si78,tr72,
 tc09,wl12,wi73,wi90,wt77,ys20,ym92}, but the question that is often ignored or eluded is: how good are these solutions?

 If the underlying boundary-value problem (BVP)is correctly formulated then the ultimate test of whether a solution is valid or not is whether it satisfies the equations inherent in the BVP, but to carry out this test is often very painstaking because it requires generating the displacement field and its gradient at all points of the ground and on the upper, curved, boundary of the protuberance, not to speak of the field everywhere within and below the protuberance. Another  manner to test the solution is by finding out if it satisfies a conservation law (such as that of energy). Since we shall be concerned with frequency-domain formulations of the BVP, the conservation of energy law takes the form of what we term the conservation of flux law which states that the input flux equals the scattered flux plus the absorbed flux \cite{wi19}.

 We show that, to employ this law in its first, abstract form, requires computing the scattering amplitude  in the far-field zone (whether the protuberance filler medium is lossy or non-lossy) as well as the displacement field at all points within the protuberance (only when the filler is lossy). The latter task (for lossy fillers) is likewise painstaking, so that it is opportune to show (as is done herein) that the said task can be replaced by computing the field and its gradient only on the lower, flat (ground-level) boundary of the protuberance. To show that the so-contrived conservation (of flux) law makes sense, we apply it to the case of a cylindrical protuberance of rectangular shape submitted to a shear-horizontal plane body wave.

 The result of this operation is that the said solution (deriving from a domain decomposition separation of variables (SOV) technique) indeed satisfies the conservation law for all orders of approximation of the solution which suggests that the SOV solution is, in a sense, exact. However,: i) if this law is not satisfied the solution cannot be reasonably qualified as exact (i.e., might even be fallacious if the difference between the input and output (scattered plus absorbed) fluxes is large, and  ii) even if this law is satisfied, the solution may be incorrect (this is demonstrated numerically herein). This means that the conservation law is a necessary, but not sufficient condition for testing the validity of a solution to the seismic response problem.
\section{Description of the seismic scattering problem}\label{desc}
In the first approximation, the earth's surface is considered to be (horizontally-) flat (termed "ground" for short) and to separate the vacuum (above) from a linear, isotropic, homogeneous (LIH) solid (below), so as to be stress-free. In the second approximation the flat ground is locally deformed so as to penetrate into what was formerly the vacuum half space. We now define the protuberance as the region between the locally-deformed stress-free surface and what was formerly a portion of the flat ground. This protuberance is underlain by the same LIH solid as previously, but the solid material within the protuberance is now assumed to be  only linear and isotropic (i.e., not homogeneous). In fact, we consider the specific  case in which the material within the protuberance is in the form of a horizontal bilayer so as to be able to account for various empirically-observed effects that are thought to be due to inhomogeneity of the protuberance material. Furthermore, we assume that: the protuberance is of infinite extent along one of the cartesian coordinates,  and its stress-free boundary to be of arbitrary shape (in its cross-section plane). The underlying problem  of much of what follows  is the prediction of the seismic wave response of this earth model.

The earthquake sources are assumed to be located in the lower half-space and to be infinitely-distant from the ground so that the seismic (pulse-like) solicitation takes the form of a body (plane) wave in the neighborhood of the protuberance. This plane wavefield is assumed to be of the shear-horizontal ($SH$) variety, which means that: only one (i.e., the cartesian coordinate $z$) component of the incident displacement field is non-nil and this field does not depend on $z$.

We shall assume that the protuberance boundary does not depend on $z$ and that the (often relatively-soft) medium filling the protuberance as well as the (usuallly relatively-hard) medium below the protuberance are both linear and isotropic. Furthermore the medium of the below-ground half space is assumed to be homogeneous, whereas that of the protuberance to be piecewise homogeneous (however, this heterogeneity is such as to not depend on $z$). It ensues that the scattered and total displacement fields within and outside the protuberance do not depend on $z$. Thus, the problem we are faced with is 2D ($z$ being the ignorable coordinate), and it is sufficient to search for the $z$-component of the scattered displacement field, designated by $u_{z}^{s}(\mathbf{x};\omega)$ in the sagittal (i.e., $x-y$) plane, when $u_{z}^{i}(\mathbf{x};\omega)$ designates the incident displacement field, with $\mathbf{x}=(x,y)$ and $\omega=2\pi f$  the angular frequency, $f$ the frequency. The temporal version of the displacement field is $u_{z}(\mathbf{x};t)=2\Re\int_{0}^{\infty}u_{z}^{i}(\mathbf{x};\omega)\exp(-i\omega t)d\omega$ wherein $t$ is the temporal variable.
\begin{figure}[ht]
\begin{center}
\includegraphics[width=0.75\textwidth]{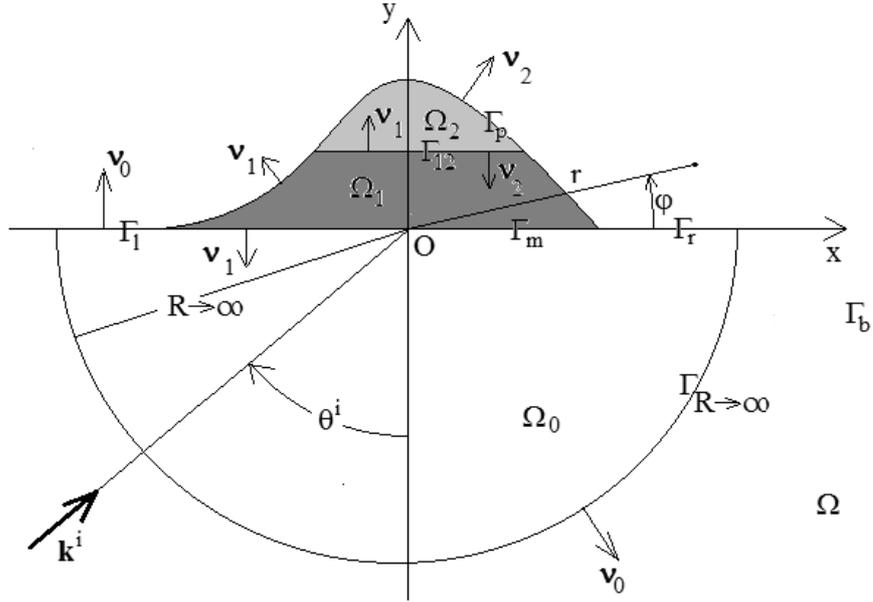}
\caption{Sagittal plane view of the 2D scattering configuration. The protuberance occupies the shaded areas and the medium within it is a horizontal bilayer.}
\label{protuberance}
\end{center}
\end{figure}

Fig. \ref{protuberance} describes the scattering configuration in the sagittal plane. In this figure, $\mathbf{k}^{i}=\mathbf{k}^{i}(\theta^{i},\omega)$ is the incident wavevector oriented so that its $z$  component is nil, and $\theta^{i}$ is the angle of incidence.

The portion of the ground outside the protuberance is stress-free but since the protuberance is assumed to be in welded contact with the surrounding below-ground medium, its lower, flat, boundary is the locus of continuous displacement and stress, as is requisite for the incident field to be able to penetrate into  the protuberance and then  be scattered outside the protuberance in the remaining lower half space.

Most of what is offered in this study does not imply any restrictions either on the shape of the stress-free portion of the boundary of the protuberance or on the (lossy or non-lossy) nature  of the medium filling the protuberance.

The three media (other than the one of the portion of the  space above the protuberance, being occupied by the vacuum, is of no interest since the field cannot penetrate therein)  are $M^{[l]}~;~l=0,1,2$  within which the real shear modulii $\mu^{[l]}~;~l=0,1,2$  and the generally-complex shear body wave velocities are $\beta^{[l]}~;~l=0,1,2$  i.e., $\beta^{[l]}=\beta^{'[l]}+i\beta^{''[l]}$, with $\beta^{'[l]}\ge 0$, $\beta^{''[l]}\le 0$, $\beta^{[l]}=\sqrt{\frac{\mu^{[l]}}{\rho^{[l]}}}$, and  $\rho^{[l]}$ the (generally-complex) mass density. The shear-wave velocity $\beta^{[0]}$ is assumed to be real, i.e., $\beta^{''[0]}=0$.
\section{Boundary-value problem}\label{bvp}
The protuberance occupies (in the sagittal plane (SP)) the finite-sized region $\Omega_{1}\bigcup\Omega_{2}$.  The  below-ground half-space occupies the region $\Omega_{0}$. $\Omega_{0}$ is entirely filled with $M^{[0]}$ whereas $\Omega_{1}$ is  filled with $M^{[1]}$ and $\Omega_{2}$ with  $M^{[2]}$.

Always in the sagittal plane, the flat ground is described by $\Gamma_{G}$, with $x,y$ the cartesian coordinates in the SP) and is composed  of three segments; $\Gamma_{l}$, $\Gamma_{m}$, and $\Gamma_{r}$, which designate the left-hand, middle, and right-hand portions respectively of $\Gamma_{G}$. The protuberance is an above-ground structure whose upper and lower boundaries (in the SP) are $\Gamma_{p}$ and $\Gamma_{m}$, the latter being of width $w$.

The analysis takes place in the space-frequency framework, so that all constitutive and field variables depend on the frequency $f$. This dependence will henceforth be implicit (e.g., $u_{z}(\mathbf{x};f)$, with $\mathbf{x}=(x,y)$, will be denoted by $u(\mathbf{x})$).

The seismic solicitation is an incident shear-horizontal (SH)  plane wave field of the form
\begin{equation}\label{1-000}
u^{i}(\mathbf{x})=a^{i}\exp(i\mathbf{k}^{i}\cdot\mathbf{x})=a^{i}\exp[i(k_{x}^{i}x+k_{z}^{i}y)]~,
\end{equation}
wherein $a^{i}=a^{i}(\omega)$ is the spectral amplitude of the seismic pulse, $\mathbf{k}^{i}=(k_{x}^{i},k_{y}^{i})$, $k_{x}^{i}=k^{[0]}\sin\theta^{i}$, $k_{y}^{i}=k^{[0]}\cos\theta^{i}$, $k^{[l]}=\omega/\beta^{[l]}~;~l=0,1,2$.

Owing to the fact that the configuration comprises three distinct regions, each in which the elastic parameters are constants as a function of the space variables, it is opportune to employ domain decomposition and (later on separation of variables). Thus, we decompose the total  field $u$ as:
\begin{equation}\label{1-010}
u(\mathbf{x})=u^{[l]}(\mathbf{x})~;~\forall\mathbf{x}\in\Omega_{l},~l=0,1,2~,
\end{equation}
with the understanding that these fields satisfy the 2D SH frequency domain elastic wave equation (i.e., Helmholtz equation)
\begin{equation}\label{1-020}
\Big(\triangle+\big(k^{[l]}\big)^{2}\Big)u^{[l]}(\mathbf{x})=0~;~\forall\mathbf{x}\in\Omega_{l},~l=0,1,2~,
\end{equation}
 with the notations $\triangle=\frac{\partial^{2}}{\partial x^{2}}+\frac{\partial^{2}}{\partial y^{2}}$ in the cartesian coordinate system of the sagittal plane.

 In addition, the field $u^{[0]}$ satisfies the radiation condition
\begin{equation}\label{1-030}
u^{[0]}(\mathbf{x})-u^{i}(\mathbf{x})-\sim \text {outgoing~wave} ~;~\|\mathbf{x}\|\rightarrow \infty~.
\end{equation}
due to the fact that $\Omega_{0}$ is unbounded (i.e., a semi-infinite domain).

The stress-free nature of the boundaries $\Gamma_{l}$, $\Gamma_{p}$, $\Gamma_{r}$, entail the boundary conditions:
\begin{equation}\label{bc-010}
\mu^{[0]}u_{,y}^{[0]}(\mathbf{x})=0~;~\forall\mathbf{x}\in\Gamma_{l}+\Gamma_{r}~,
\end{equation}
\begin{equation}\label{bc-020}
\mu^{[2]}u_{,y}^{[2]}(\mathbf{x})=0~;~\forall\mathbf{x}\in\Gamma_{p}~,
\end{equation}
wherein   $u_{,\zeta}$ denotes the first  partial derivative of $u$ with respect to $\zeta$.

The fact, that the horizontal segment $\Gamma_{12}$ between the two media filling the protuberance is assumed to be an interface across which two media are in welded contact, entails the continuity conditions:
\begin{equation}\label{bc-040}
u^{[2]}(\mathbf{x})-u^{[1]}(\mathbf{x})=0~;~\forall\mathbf{x}\in\Gamma_{12}~,
\end{equation}
\begin{equation}\label{bc-050}
\mu^{[2]}u_{,y}^{[2]}(\mathbf{x})-\mu^{[1]}u_{,y}^{[1]}(\mathbf{x})=0~;~\forall\mathbf{x}\in\Gamma_{12}~.
\end{equation}

Finally, the fact, that $\Gamma_{m}$ was assumed to be an interface across which two media are in welded contact, entails the continuity conditions:
\begin{equation}\label{bc-040}
u^{[0]}(\mathbf{x})-u^{[1]}(\mathbf{x})=0~;~\forall\mathbf{x}\in\Gamma_{m}~,
\end{equation}
\begin{equation}\label{bc-050}
\mu^{[0]}u_{,y}^{[0]}(\mathbf{x})-\mu^{[1]}u_{,y}^{[1]}(\mathbf{x})=0~;~\forall\mathbf{x}\in\Gamma_{m}~,
\end{equation}

The purpose of addressing such a boundary-value (direct) problem is to determine $u^{[l]}(\mathbf{x});~l=0,1,2$ for various solicitations  and parameters  relative to the  geometries of, and media filling, $\Omega_{l}~;~l=0,1,2$. The principal ambition of the analysis which follows is rather  to establish a conservation law governing $u^{[l]}(\mathbf{x})~;~l=0,1$, this being done, in the first part of our study, without actually solving for $u^{[l]}(\mathbf{x})~;~l=0,1$. However, to show that this conservation law makes sense and is useful, we shall, in the second part of this study, appeal to a separation-of-variables (SOV) solution of the problem in which the protuberance is of rectangular shape.
\section{Basic ingredients of the conservation of flux law}\label{bicl}
Eq. (\ref{1-020}) yields
\begin{equation}\label{2-020}
\Big(\triangle+\big[\big(k^{[l]}\big)^{2}\big]^{*}\Big)u^{[l]*}(\mathbf{x})=0~;~\forall\mathbf{x}\in\Omega_{l},~l=0,1,2~,
\end{equation}
wherein $(X+iY)^{*}=X-iY$. It follows (with $d\Omega$ the differential surface element) that
\begin{equation}\label{2-030}
\int_{\Omega_{l}}u^{[l]*}\mathbf{x})\Big\{\Big(\triangle+\big(k^{[l]}\big)^{2}\Big)u^{[l]}(\mathbf{x})-
u^{[l]*}(\mathbf{x})\Big(\triangle+\big[\big(k^{[l]}\big)^{2}\big]^{*}\Big)u^{[l]}(\mathbf{x})\Big\}d\Omega
=0~;~\forall\mathbf{x}\in\Omega_{l},~l=0,1,2~,
\end{equation}
or
\begin{multline}\label{2-040}
\int_{\Omega_{l}}\Big\{u^{[l]*}(\mathbf{x})\triangle u^{[l]}(\mathbf{x})-u^{[l]}(\mathbf{x})\triangle u^{*[l]}(\mathbf{x})\Big\}d\Omega+\\
\int_{\Omega_{l}}\Big\{\big(k^{[l]}\big)^{2}-\big[\big(k^{[l]}\big)^{2}\big]^{*}\Big\}\|u^{[l]}(\mathbf{x})\|^{2}d\Omega
=0~;~\forall\mathbf{x}\in\Omega_{l},~l=0,1,2~.
\end{multline}
We want to apply Green's second identity to the first integral, and to do this we must define the (closed) boundaries of $\Omega_{l}$. We already know that the boundaries of $\Omega_{1}~;~l=1,2$ are closed, but until now, $\Omega_{0}$ was not closed. To close it, we imagine a semicircle $\Gamma_{\mathcal{R}}$, of large radius $\mathcal{R}$ (taken to be infinitely large in the limit), centered at the origin $O$,  to be drawn so as to intersect the ground at $x=\pm \mathcal{R}$ and to intersect the $y$ axis at $y=-\mathcal{R}$. Thus, designating by $\Gamma_{p}^{\pm}$  the upper(lower) portions of $\Gamma_{p}$, the closed boundaries of $\Omega_{l}~;~l=0,1,2$ are:
\begin{equation}\label{2-050}
\partial_{\Omega_{2}}=\Gamma_{p}^{+}\cup\Gamma_{12}~~,~~\partial_{\Omega_{1}}=\Gamma_{m}\cup\Gamma_{p}^{-}\cup\Gamma_{12}~~,~~
\partial_{\Omega_{0}}=\Gamma_{l}\cup\Gamma_{b}\cup\Gamma_{r}\cup\Gamma_{\mathcal{R}\rightarrow\infty}~,
\end{equation}
and we shall designate by $\boldsymbol{\nu}_{l}$ the unit vector normal to $\partial\Omega_{l}$  that points towards the exterior of $\partial\Omega_{l}$. We now apply Green's second identity to obtain (with $d\Gamma$ the differential arc element)
\begin{multline}\label{2-060}
\int_{\partial\Omega_{l}}\Big\{u^{[l]*}(\mathbf{x})\boldsymbol{\nu}_{l}\cdot\nabla u^{[l]}(\mathbf{x})-u^{[l]}(\mathbf{x}) \boldsymbol{\nu}_{l}\cdot\nabla u^{*[l]}(\mathbf{x})\Big\}d\Omega+\\
\int_{\Omega_{l}}\Big\{\big(k^{[l]}\big)^{2}-\big[\big(k^{[l]}\big)^{2}\big]^{*}\Big\}\|u^{[l]}(\mathbf{x})\|^{2}d\Omega
=0~;~\forall\mathbf{x}\in\Omega_{l},~l=0,1,2~.
\end{multline}
or, in condensed form (on account of the non-lossy nature of the solid filling $\Omega_{0}$ and the homogeneous nature of the solids filling $\Omega_{l}~;~l=1,2$):
\begin{equation}\label{2-070}
\Im\int_{\partial\Omega_{0}}u^{[0]*}(\mathbf{x})\boldsymbol{\nu}_{0}\cdot\nabla u^{[0]}(\mathbf{x})d\Gamma=0~,
\end{equation}
\begin{equation}\label{2-080}
\Im\int_{\partial\Omega_{l}}u^{[l]*}(\mathbf{x})\boldsymbol{\nu}_{l}\cdot\nabla u^{[l]}(\mathbf{x})d\Gamma-\Im\big[\big(k^{[l]}\big)^{2}\big]\int_{\Omega_{1}}\|u^{[l]}(\mathbf{x})\|^{2}d\Omega=0~;~l=1,2~.
\end{equation}
More explicitly, and on account of (\ref{2-050}),
\begin{multline}\label{2-090}
\Im\int_{\Gamma_{l}+\Gamma_{r}}u^{[0]*}(\mathbf{x})\boldsymbol{\nu}_{0}\cdot\nabla u^{[0]}(\mathbf{x})d\Gamma+
\Im\int_{\Gamma_{m}}u^{[0]*}(\mathbf{x})\boldsymbol{\nu}_{0}\cdot\nabla u^{[0]}(\mathbf{x})d\Gamma+\\
\Im\int_{\Gamma_{\infty}}u^{[0]*}(\mathbf{x})\boldsymbol{\nu}_{0}\cdot\nabla u^{[0]}(\mathbf{x})d\Gamma=0~,
\end{multline}
\begin{multline}\label{2-100}
\Im\int_{\Gamma_{m}}u^{[1]*}(\mathbf{x})\boldsymbol{\nu}_{1}\cdot\nabla u^{[1]}(\mathbf{x})d\Gamma+
\Im\int_{\Gamma_{12}}u^{[1]*}(\mathbf{x})\boldsymbol{\nu}_{1}\cdot\nabla+
\Im\int_{\Gamma_{p}^{-}}u^{[1]*}(\mathbf{x})\boldsymbol{\nu}_{1}\cdot\nabla u^{[1]}(\mathbf{x})d\Gamma u^{[1]}(\mathbf{x})d\Gamma+\\
\Im\big[\big(k^{[1]}\big)^{2}\big]\int_{\Omega_{1}}\|u^{[1]}(\mathbf{x})\|^{2}d\Omega=0~,
\end{multline}
\begin{equation}\label{2-105}
\Im\int_{\Gamma_{12}}u^{[2]*}(\mathbf{x})\boldsymbol{\nu}_{2}\cdot\nabla u^{[1]}(\mathbf{x})d\Gamma+
\Im\int_{\Gamma_{p}}u^{[2]*}(\mathbf{x})\boldsymbol{\nu}_{2}\cdot\nabla u^{[2]}(\mathbf{x})d\Gamma+\Im\big[\big(k^{[2]}\big)^{2}\big]\int_{\Omega_{2}}\|u^{[2]}(\mathbf{x})\|^{2}d\Omega=0~.
\end{equation}
Due to  $\boldsymbol{\nu}_{2}(\mathbf{x})=-\boldsymbol{\nu}_{1}(\mathbf{x})~;~\forall\mathbf{x}\in\Gamma_{b}$,
and the boundary and continuity conditions, we have:
\begin{equation}\label{2-110}
u^{[2]*}(\mathbf{x})\boldsymbol{\nu}_{2}\cdot\nabla u^{[2]}(\mathbf{x})=
-\frac{\mu^{[1]}}{\mu^{[2]}}u^{[1]*}(\mathbf{x})\boldsymbol{\nu}_{1}\cdot\nabla u^{[1]}(\mathbf{x})~;~\forall\mathbf{x}\in\Gamma_{12}~,
\end{equation}
so that, since $\frac{\mu^{[1]}}{\mu^{[2]}}$ was assumed to be real,  (\ref{2-105}) becomes:
\begin{equation}\label{2-115}
-\frac{\mu^{[1]}}{\mu^{[2]}}\Im\int_{\Gamma_{12}}u^{[1]*}(\mathbf{x})\boldsymbol{\nu}_{1}\cdot\nabla u^{[1]}(\mathbf{x})d\Gamma+
\Im\big[\big(k^{[2]}\big)^{2}\big]\int_{\Omega_{2}}\|u^{[2]}(\mathbf{x})\|^{2}d\Omega=0~.
\end{equation}
Due to the stress-free boundary condition, (\ref{2-100}) reduces to
\begin{equation}\label{2-120}
\Im\int_{\Gamma_{m}}u^{[1]*}(\mathbf{x})\boldsymbol{\nu}_{1}\cdot\nabla u^{[1]}(\mathbf{x})d\Gamma+
\Im\int_{\Gamma_{12}}u^{[1]*}(\mathbf{x})\boldsymbol{\nu}_{1}\cdot\nabla+
\Im\big[\big(k^{[1]}\big)^{2}\big]\int_{\Omega_{1}}\|u^{[1]}(\mathbf{x})\|^{2}d\Omega=0~,
\end{equation}
so that the linear combination of these last two equations gives rise to:
\begin{equation}\label{2-125}
\Im\int_{\Gamma_{m}}u^{[1]*}(\mathbf{x})\boldsymbol{\nu}_{1}\cdot\nabla u^{[1]}(\mathbf{x})d\Gamma+
\frac{\mu^{[2]}}{\mu^{[1]}}\Im\big[\big(k^{[2]}\big)^{2}\big]\int_{\Omega_{2}}\|u^{[2]}(\mathbf{x})\|^{2}d\Omega+
\Im\big[\big(k^{[1]}\big)^{2}\big]\int_{\Omega_{1}}\|u^{[1]}(\mathbf{x})\|^{2}d\Omega=0~.
\end{equation}
Due to  $\boldsymbol{\nu}_{0}(\mathbf{x})=-\boldsymbol{\nu}_{1}(\mathbf{x})~;~\forall\mathbf{x}\in\Gamma_{b}$,
and the boundary and continuity conditions, we have:
\begin{equation}\label{2-130}
u^{[0]*}(\mathbf{x})\boldsymbol{\nu}_{0}\cdot\nabla u^{[0]}(\mathbf{x})=
-\frac{\mu^{[1]}}{\mu^{[0]}}u^{[1]*}(\mathbf{x})\boldsymbol{\nu}_{1}\cdot\nabla u^{[1]}(\mathbf{x})~;~\forall\mathbf{x}\in\Gamma_{m}~,
\end{equation}
\begin{equation}\label{2-135}
u^{[0]*}(\mathbf{x})\boldsymbol{\nu}_{0}\cdot\nabla u^{[0]}(\mathbf{x})=0~;~\forall\mathbf{x}\in\Gamma_{l}+\Gamma_{r}~,
\end{equation}
so that (\ref{2-090}) becomes
\begin{equation}\label{2-140}
-\frac{\mu^{[1]}}{\mu^{[0]}}\Im\int_{\Gamma_{m}}u^{[1]*}(\mathbf{x})\boldsymbol{\nu}_{1}\cdot\nabla u^{[1]}(\mathbf{x})+\\
\Im\int_{\Gamma_{\infty}}u^{[0]*}(\mathbf{x})\boldsymbol{\nu}_{0}\cdot\nabla u^{[0]}(\mathbf{x})d\Gamma=0~,
\end{equation}
whence the linear combination of (\ref{2-090}) and (\ref{2-125}) gives rise to:
\begin{equation}\label{2-150}
\Im\int_{\Gamma_{\infty}}u^{[0]*}(\mathbf{x})\boldsymbol{\nu}_{0}\cdot\nabla u^{[0]}(\mathbf{x})d\Gamma+
\frac{\mu^{[2]}}{\mu^{[0]}}\Im\big[\big(k^{[2]}\big)^{2}\big]\int_{\Omega_{2}}\|u^{[2]}(\mathbf{x})\|^{2}d\Omega+
\frac{\mu^{[1]}}{\mu^{[0]}}\Im\big[\big(k^{[1]}\big)^{2}\big]\int_{\Omega_{1}}\|u^{[1]}(\mathbf{x})\|^{2}d\Omega=0~.
\end{equation}
Eqs. (\ref{2-140}) and (\ref{2-150}) are  alternate expressions of the same sought-for conservation law.
\section{Separation-of-variables representation of the field in the half space underneath the protuberance}
The fact that the conservation law involves the field on $\Gamma_{\infty}=\Gamma_{\mathcal{R}\rightarrow\infty}$ means that we must dispose of an expression for the field $u^{[0]}(\mathbf{x})$ in the half space (and notably in the far-field zone thereof) underneath the protuberance. To do this, we are not obliged to solve the forward scattering problem, but only obtain a {\it representation} of $u^{[0]}(\mathbf{x})$ that obeys the Helmholtz equation, the radiation condition and the stress-free boundary condition on $\Gamma_{l}+\Gamma_{r}$. To do this, we employ the separation-of-variables (SOV) technique in terms of the cartesian coordinates to obtain
\begin{equation}\label{3-010}
u^{s}=u^{[0]}(\mathbf{x})-u^{i}(\mathbf{x})-u^{r}(\mathbf{x})=\int_{-\infty}^{\infty}\mathcal{B}(k_{x})
\exp[i(k_{x}x-k^{[0]}_{y}y)]\frac{dk_{x}}{k^{[0]}_{y}}~;~\forall y\le 0~,~\forall x\in\mathbb{R}~,
\end{equation}
wherein
\begin{equation}\label{3-020}
k^{[0]}_{y}=\sqrt{\big(k^{[0]}\big)^{2}-\big(k_{x}\big)^{2}}
~~,~~\Re k^{[0]}_{y}\ge 0~;~\forall\omega\ge 0~~,~~
\Im k^{[0]}_{y}\ge 0~;~\forall\omega\ge 0~.
\end{equation}
and
\begin{equation}\label{3-030}
u^{r}(x,y)=u^{i}(x,-y)~.
\end{equation}
It follows that:
\begin{equation}\label{3-040}
u^{i}(x,0)+u^{r}(x,0)=2u^{i}(x,0)=2a^{i}\exp[ik_{x}x]~,
\end{equation}
\begin{equation}\label{3-050}
u_{,y}^{i}(x,0)+u_{,y}^{r}(x,0)=0~,
\end{equation}
whence
\begin{equation}\label{3-060}
u^{[0]}(x,0)=2u^{i}(x,0)+u^{s}(x,0)~,
\end{equation}
\begin{equation}\label{3-070}
u_{,y}^{[0]}(x,0)=u_{,y}^{s}(x,0)=-i\int_{-\infty}^{\infty}\mathcal{B}(k_{x})\exp[i(k_{x}x]~.
\end{equation}
Fourier inversion then yields
\begin{equation}\label{3-080}
\mathcal{B}(k_{x})=\frac{i}{2\pi}\int_{-\infty}^{\infty}u_{,y'}^{[0]}(x',0)\exp[-i(k_{x}x']dx'~.
\end{equation}
We assume that $\Gamma_{m}$, whose width is $w$, extends from $x'=-w/2$ to $x'=w/2$, and make  use of the stress-free boundary condition on $\Gamma_{l}+\Gamma_{r}$ to obtain, with the change of variables $k_{x}=k^{[0]}\cos\phi$:
\begin{equation}\label{3-085}
B(\phi)=\pi\mathcal{B}(k_{x})=\frac{i}{2\pi}\int_{-w/2}^{w/2}u_{,y'}^{[0]}(x',0)\exp[-ik^{[0]}x'\cos\phi]dx~.
\end{equation}
The introduction of (\ref{3-080}) into (\ref{3-010}) gives, after the interchange of the orders of integration
\begin{equation}\label{3-090}
u^{s}(\mathbf{x})=\frac{i}{2\pi}\int_{-w/2}^{w/2}u_{,y'}^{[0]}(x',0)H_{0}^{(1)}(k^{[0]}\mathcal{R})dx'~;~\forall y\le 0~,~\forall x\in\mathbb{R}~.
\end{equation}
wherein $\mathcal{R}=\sqrt{(x-x')^{2}+y^{2}}$ and
\begin{equation}\label{3-100}
H_{0}^{(1)}(k^{[0]}\mathcal{R})=\frac{1}{\pi}\int_{-\infty}^{\infty}\exp[ik_{x}(x-x')-k_{y}y)]\frac{dk_{x}}{k^{[0]}_{y}}~;~\forall y\le 0~,~\forall x\in\mathbb{R}~
\end{equation}
is the zeroth-order Hankel function of the first kind \cite{as68}.

Since we are particularly interested in the field in the far-field zone, we appeal to the well-known \cite{as68} asymptotic form of the Hankel function
\begin{equation}\label{3-110}
H_{0}^{(1)}\big(k^{[0]}\|\mathbf{x}-(x',0)\|\big)=H_{0}^{(1)}\big(k^{[0]}\mathcal{R}\big)\sim\Big(\frac{2}{\pi k^{[0]}\mathcal{R}}\Big)^{1/2}\exp[i(k^{[0]}\mathcal{R}-\pi/4)]~;~k^{[0]}\mathcal{R}\rightarrow\infty~.
\end{equation}
The change of variables $x=r\cos\phi$, $y=r\cos\phi$, $x'=r\cos\phi'$ (with the understanding that $\phi'=0$ for $x>0$ and $\phi'=\pi$ for $x<0$) gives rise to
\begin{equation}\label{3-120}
\mathcal{R}=\sqrt{((r\cos\phi-r'\cos\phi')^{2}+(r\sin\phi)^{2}}\sim r-x'\cos\phi~;~\frac{r'}{r}<<1~,
\end{equation}
the condition $\frac{r'}{r}<<1$ being verified when $|x'|=r'\le w/2$ and $r\rightarrow\infty$. Consequently,
\begin{equation}\label{3-130}
u^{s}(\mathbf{x})\sim B(\phi)\Big(\frac{2}{\pi k^{[0]}r}\Big)^{1/2}\exp[i(k^{[0]}r-\pi/4)]~;~|x'|=r'\le w/2~~,~~k^{[0]}r\rightarrow\infty~.
\end{equation}
wherein $B(\phi)$ is as in (\ref{3-085}).

What this all means is that the scattered field behaves asymptotically (i.e., for  $k^{[0]}r\rightarrow\infty$) like a cylindrical wave whose complex amplitude is $B(\phi)$. We now dispose of the means for evaluating the integral on $\Gamma_{\infty}$ in the conservation of flux relations.
\section{Incorporation of the asymptotic form of the field in the conservation law}
Now let us return to (\ref{2-140})-(\ref{2-150}) which  can be written either  as
\begin{equation}\label{3-400}
I-J=0~.
\end{equation}
or
\begin{equation}\label{3-405}
I+K=0~.
\end{equation}
with
\begin{equation}\label{3-410}
I=\Im\int_{\Gamma_{\infty}}u^{[0]*}(\mathbf{x})\boldsymbol{\nu}_{0}\cdot\nabla u^{[0]}(\mathbf{x})d\Gamma~,
\end{equation}
and let us examine $I$ more closely. On account of (\ref{3-060}) we have
\begin{equation}\label{3-420}
I=\Im\int_{\Gamma_{\infty}}[u^{*}_{i}+u^{*}_{r}+u^{*}_{s}]\boldsymbol{\nu}_{0}\cdot\nabla [u_{i}+u_{r}+u_{s}]d\Gamma=
(I^{ii}+I^{rr})+(I^{ir}+I^{ri})+(I^{is}+I^{si})+(I^{rs}+I^{sr})+I^{ss}~,
\end{equation}
or, due to the facts that:
                                 $d\Gamma\big|_{\Gamma_{\mathcal{R}}}=\mathcal{R}d\phi$ and
$\boldsymbol{\nu}_{0}\cdot\nabla u^{[0]}\big|_{\Gamma_{\mathcal{R}}}=
u_{,r}^{[0]}(\mathcal{R},\phi)$,
we obtain
\begin{equation}\label{3-430}
I=\Im~\lim_{\mathcal{R}\rightarrow\infty}\int_{\pi}^{2\pi}[u^{i*}(\mathcal{R},\phi)+u^{r*}(\mathcal{R},\phi)+u^{s*}(\mathcal{R},\phi)]
[
u_{,r}^{i}(\mathcal{R},\phi)+
u_{,r}^{r}(\mathcal{R},\phi)+
u_{,r}^{s}(\mathcal{R},\phi)
]
\mathcal{R}d\phi~,
\end{equation}
in which we must adopt the following cylindrical coordinate representations of $u^{i}$ and $u^{r}$, wherein $x=r\cos\phi$ and $y=r\sin\phi$:
\begin{equation}\label{3-440}
  u^{i}=a^{i}\exp[ir(k_{x}^{i}\cos\phi+k_{y}^{i}\sin\phi)]~~,
~~u^{r}=a^{i}\exp[ir(k_{x}^{i}\cos\phi-k_{y}^{i}\sin\phi)]~.
\end{equation}
It is easy to show  \cite{wi19} that:
\begin{equation}\label{3-450}
I^{ii}+I^{rr}=I^{ir}+I^{ri}=I^{is}+I^{si}=0~.
\end{equation}
In \cite{wi19} we showed that the employment of the stationary phase technique enables  to find:
\begin{equation}\label{3-480}
I^{rs}+I^{sr}=4\Re~\big[a^{i*}B(3\pi/2+\theta^{i})\big]~.
\end{equation}
The employment of (\ref{3-130}) yields (see \cite{wi19})
\begin{equation}\label{3-495}
I^{ss}=\frac{2}{\pi}\int_{\pi}^{2\pi}\|B(\phi)\|^{2}d\phi~.
\end{equation}
The last step is to carry out the sum in (\ref{3-420} so as to obtain
\begin{equation}\label{3-500}
I=\frac{2}{\pi}\int_{\pi}^{2\pi}\|B(\phi)\|^{2}d\phi+4\Re~\big[a^{i*}B(3\pi/2+\theta^{i})\big]~,
\end{equation}
which is the detailed expression of the term $I$ in the conservation law which we wrote either as $I-J=0$ or $I+K$.
\section{Explicit forms and interpretation of the conservation of flux relation}
Let us now examine in detail the term $J$ of this law. The definition of $J$ is
\begin{equation}\label{3-505}
J=\frac{\mu^{[1]}}{\mu^{[0]}}\Im\int_{\Gamma_{b}}u^{[1]*}(\mathbf{x})\boldsymbol{\nu}_{1}\cdot\nabla u^{[1]}(\mathbf{x})d\Gamma~,
\end{equation}
but, as underlined earlier, our ambition was not to solve the boundary-value problem, namely for the field $u^{[1]}(\mathbf{x})$ within the basin, so that we cannot go beyond expressing the conservation law as
\begin{equation}\label{3-507}
I-J=\frac{2}{\pi}\int_{\pi}^{2\pi}\|B(\phi\|^{2}d\phi+4\Re~\big[a^{i*}B(3\pi/2+\theta^{i})\big]-
\frac{\mu^{[1]}}{\mu^{[0]}}\Im\int_{\Gamma_{b}}u^{[1]*}(\mathbf{x})\boldsymbol{\nu}_{1}\cdot\nabla u^{[1]}(\mathbf{x})d\Gamma=0~,
\end{equation}
whose meaning, is unfortunately not clear for the moment.

To cope with this problem, we make use of the alternative expression (\ref{2-140}) of the conservation law
\begin{equation}\label{2-510}
I+K=\Im\int_{\Gamma_{\infty}}u^{[0]*}(\mathbf{x})\boldsymbol{\nu}_{0}\cdot\nabla u^{[0]}(\mathbf{x})d\Gamma+
\sum_{j=1}^{2}\frac{\mu^{[j]}}{\mu^{[0]}}\Im\big[\big(k^{[j]}\big)^{2}\big]\int_{\Omega_{j}}\|u^{[j]}(\mathbf{x})\|^{2}d\Omega=0~,
\end{equation}
which authorizes us to write (\ref{2-150}) as
\begin{equation}\label{3-520}
I+K=\frac{2}{\pi}\int_{\pi}^{2\pi}\|B(\phi)\|^{2}d\phi+4\Re~\big[a^{i*}B(3\pi/2+\theta^{i})\big]+
\sum_{j=1}^{2}\frac{\mu^{[j]}}{\mu^{[0]}}\Im\big[\big(k^{[j]}\big)^{2}\big]\int_{\Omega_{j}}\|u^{[j]}(\mathbf{x})\|^{2}d\Omega=0~.
\end{equation}
 This expression informs us that if the media within the ~~protuberance domain ~$\Omega_{1}\cup\Omega_{2}$ are ~~non-lossy, ~then~ $\Im k^{[1]}=\Im k^{[2]}=0$ so~ that  $K=\sum_{j=1}^{2}\frac{\mu^{[j]}}{\mu^{[0]}}\Im\big[\big(k^{[j]}\big)^{2}\big]\int_{\Omega_{j}}\|u^{[j]}(\mathbf{x})\|^{2}d\Omega=0$,~~ which~~ entails
$J=\frac{\mu^{[1]}}{\mu^{[0]}}\Im\int_{\Gamma_{b}}u^{[1]*}(\mathbf{x})\boldsymbol{\nu}_{1}\cdot\nabla u^{[1]}(\mathbf{x})d\Gamma=0$. This means that $J$ accounts for damping (i.e., absorption) in $\Omega_{1}$ since $J$ vanishes in the absence of a loss mechanism, as expressed by $c^{[j]}~;~j=1,2$ (and therefore $k^{[j]}~;~j=1,2$) being complex. A valid question is then: damping/absorption of what? It is not energy because the units of $J=-K$ are not those of energy, but rather something related to energy, which we shall name 'flux'. If we reason in terms of energy, the only means by which energy can be lost in a scattering problem such as ours is by radiation damping, which is the mechanism by which (scattered) energy escapes to the outer reaches of $\Omega_{0}$, i.e., escapes to $r\rightarrow\infty$ in the half-space $y<0$ due to the fact that this half-space is not bounded for negative $y$. As stated earlier, radiation damping is exclusively related to the scattered wave portion of the total field, and the function that expresses this relation to the scattered field is $B(\phi)$, so that the the term expressing radiation damping must be $\frac{2}{\pi}\int_{\pi}^{2\pi}\|B(\phi)\|^{2}d\phi$ in $I$. We are therefore authorized to call this term the 'scattered flux'. Now to continue to reason in terms of energy, it is pertinent to ask: if 'lost energy',  is the sum of absorbed and radiation damping energies, then what is the 'provided energy', if we accept the fact that energy must be conserved, i.e.,  'lost energy'='provided energy'? In our problem, the means by which energy is provided is obviously via the incident plane wave, or in other terms: the means by which flux is provided is via the incident plane wave and the only term in our 'conservation of flux' relation that can account for this is $4\Re~\big[a^{i*}B(3\pi/2+\theta^{i})\big]$ in $I$. Since this term also contains a quantity related to the scattered field (via $B(3\pi/2+\theta^{i})$ ) it is more coherent to normalize the expression of the conservation of flux law $I-J=0$ by dividing it by $-4\Re~\big[a^{i*}B(3\pi/2+\theta^{i})\big]$ so as to obtain
\begin{equation}\label{3-530}
\frac{\frac{-2}{\pi}\int_{\pi}^{2\pi}\|B(\phi)\|^{2}d\phi}{4\Re~\big[a^{i*}B(3\pi/2+\theta^{i})\big]}+
\frac{\frac{\mu^{[1]}}{\mu^{[0]}}\Im\int_{\Gamma_{b}}u^{[1]*}(\mathbf{x})\boldsymbol{\nu}_{1}\cdot\nabla u^{[1]}(\mathbf{x})d\Gamma}
{4\Re~\big[a^{i*}B(3\pi/2+\theta^{i})\big]}=1~,
\end{equation}
which can be written as
\begin{equation}\label{3-540}
\mathcal{S}+\mathcal{A}=\mathcal{I}~,
\end{equation}
wherein
\begin{equation}\label{3-550}
\mathcal{S}=\frac{-\int_{\pi}^{2\pi}\|B(\phi)\|^{2}\frac{d\phi}{\pi}}
{2\Re~\big[a^{i*}B(3\pi/2+\theta^{i})\big]}=\frac{\mathcal{-J}}{\mathcal{K}}~,
\end{equation}
is the so-called 'normalized scattered flux',
\begin{equation}\label{3-560}
\mathcal{A}=\frac{\frac{\mu^{[1]}}{2\mu^{[0]}}\Im\int_{\Gamma_{b}}u^{[1]*}(\mathbf{x})\boldsymbol{\nu}_{1}\cdot\nabla u^{[1]}(\mathbf{x})d\Gamma}
{2\Re~\big[a^{i*}B(3\pi/2+\theta^{i})\big]}=\frac{\mathcal{L}}{\mathcal{K}}~,
\end{equation}
is the so-called 'normalized absorbed flux', and
\begin{equation}\label{3-570}
\mathcal{I}=1~.
\end{equation}
is the so-called 'normalized incident flux'. In black-box language, we can say the the conservation law expresses the fact that the 'input flux' $\mathcal{I}$ equals the 'output flux' $\mathcal{S}+\mathcal{A}$, the latter being the sum of the scattered flux $\mathcal{S}$ and the 'absorbed flux' $\mathcal{A}$, wherein, for convenience we have dropped the term 'normalized' which is henceforth implicit.

Note that (\ref{3-540})-(\ref{3-570}) are in agreement with the conservation of flux relation previously obtained in \cite{wi19} for the case of a  basin filled with a lossy or non-lossy medium.

Note also that in the case the protuberance media are non-lossy, the conservation law does not depend explicitly on any of the constitutive parameters of the filler, i.e.,  $c^{[j]}=c^{'[j]}~;~j=1,2$ and $\mu^{[j]}~;~j=1,2$. However it does depend implicitly on these parameters via $B(\phi)$.
\section{Demonstration that the conservation of flux relation is satisfied by the  formal solution of the scattering problem when the protuberance is of rectangular shape and is composed of two lossy or lossless media}
%
\subsection{Description of the configuration}
\begin{figure}[ht]
\begin{center}
\includegraphics[width=0.85\textwidth]{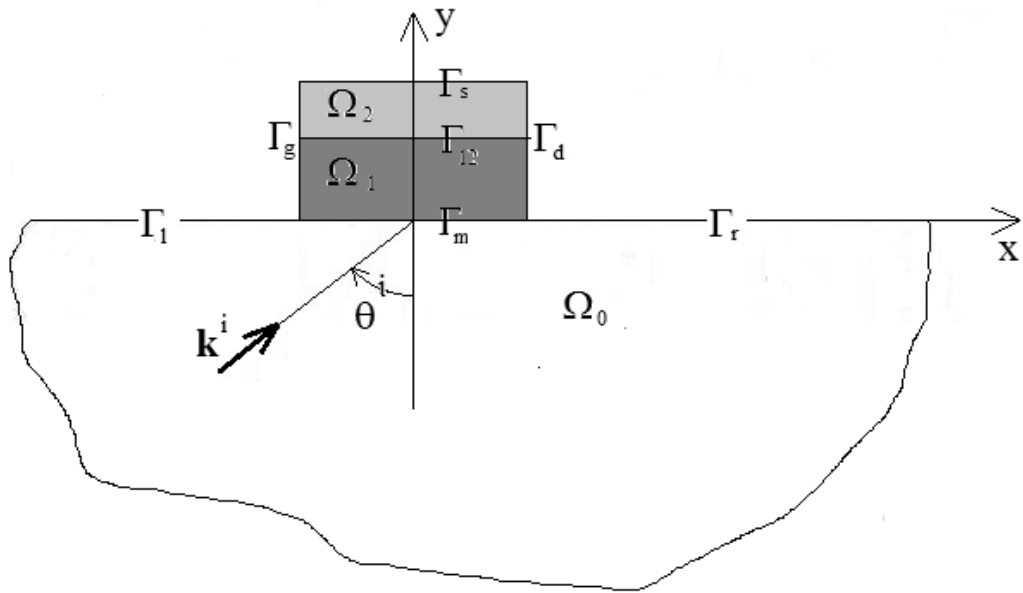}
\caption{Sagittal plane view of the 2D rectangular protuberance scattering configuration. Note that now the boundary $\Gamma_{p}$ of the above-ground feature is composed of three connected portions, $\Gamma_{g}$, $\Gamma_{s}$ and $\Gamma_{d}$.}
\label{hill}
\end{center}
\end{figure}

From now on, the option is to completely solve the forward scattering problem.
Fig.\ref{hill} describes the scattering configuration in which the bilayer protuberance outer boundary (in the sagittal plane) is  rectangular.
As previously, the width of the protuberance is $w$, and its other characteristic dimensions are the bottom ($h_{1}$) and top ($h_{2}$) layer thicknesses, with $h=h_{1}+h_{2}$ being the height of the protuberance. What was formerly $\Gamma_{p}$ is now $\Gamma_{g}\cup\Gamma_{s}\cup\Gamma_{d}$, wherein $\Gamma_{g}$ is the leftmost vertical segment of height $h$, $\Gamma_{s}$ is the top segment of width $w$ and $\Gamma_{d}$ is the rightmost vertical segment of height $h$. Everything else is as in fig.\ref{protuberance}.
\subsection{Boundary-value problem}\label{bvp}
Owing to the fact that the configuration comprises three distinct regions, each in which the elastic parameters are constants as a function of the space variables, it is opportune to employ domain decomposition and separation of variables (DD-SOV). Thus, as previously, we decompose the total  field $u$ as:
\begin{equation}\label{5-005}
u(\mathbf{x})=u^{[l]}(\mathbf{x})~;~\forall\mathbf{x}\in\Omega_{l},~l=0,1,2~,
\end{equation}
with the understanding that these fields satisfy the 2D SH frequency domain elastic wave equation (i.e., Helmholtz equation)
and $u^{[0]}$ satisfies the radiation condition

The stress-free nature of the boundaries $\Gamma_{g}$, $\Gamma_{d}$, $\Gamma_{s}$, $\Gamma_{l}$ and $\Gamma_{r}$ entail the boundary conditions:
\begin{equation}\label{5-010}
\mu^{[0]}u_{,y}^{[0]}(\mathbf{x})=0~;~\forall\mathbf{x}\in\Gamma_{l}+\Gamma_{r}~,
\end{equation}
\begin{equation}\label{5-020}
\mu^{[l]}u^{[l]}_{,x}(\mathbf{x})=0~;~\forall\mathbf{x}\in\Gamma_{g}+\Gamma_{d}~,~l=1,2~,
\end{equation}
\begin{equation}\label{5-030}
\mu^{[1]}u_{,y}^{[1]}(\mathbf{x})=0~;~\forall\mathbf{x}\in\Gamma_{s}~.
\end{equation}

Finally, the fact that $\Gamma_{12}$ and $\Gamma_{m}$ were assumed to be interfaces across which two media are in welded contact, entails the continuity conditions:
\begin{equation}\label{5-040}
u^{[0]}(\mathbf{x})-u^{[1]}(\mathbf{x})=0~;~\forall\mathbf{x}\in\Gamma_{m}~,
\end{equation}
\begin{equation}\label{5-050}
\mu^{[0]}u_{,y}^{[0]}(\mathbf{x})-\mu^{[1]}u_{,y}^{[1]}(\mathbf{x})=0~;~\forall\mathbf{x}\in\Gamma_{m}~,
\end{equation}
\begin{equation}\label{5-060}
u^{[1]}(\mathbf{x})-u^{[2]}(\mathbf{x})=0~;~\forall\mathbf{x}\in\Gamma_{12}~,
\end{equation}
\begin{equation}\label{5-070}
\mu^{[1]}u_{,y}^{[1]}(\mathbf{x})-\mu^{[2]}u_{,y}^{[2]}(\mathbf{x})=0~;~\forall\mathbf{x}\in\Gamma_{12}~.
\end{equation}
%
\subsection{Field representations via separation of variables (SOV)}\label{sov}
 As in the case of arbitrarily-shaped protuberances,  the SOV technique gives rise to the field representation
\begin{equation}\label{6-010}
u^{[0]}(\mathbf{x})=u^{i}(\mathbf{x})+u^{r}(\mathbf{x})+u^{s}(\mathbf{x})
\end{equation}
wherein (in a somewhat simpler notation)
\begin{equation}\label{6-014}
u^{s}(\mathbf{x})=\int_{-\infty}^{\infty}\mathcal{B}(k_{x})f(k_{x},x)\exp(-ik_{y}y)\frac{dk_{x}}{k^{[0]}_{y}}~,
\end{equation}
with:
\begin{equation}\label{6-013}
f(k_{x},x)=\exp(ik_{x}x)~.
\end{equation}

Note that the scattered field $u^{s}$ is expressed as a sum of plane waves, some of which are propagative (for real $k^{[0]}_{y}$) and the others evanescent (for imaginary $k^{[0]}_{y}$).

Within the  rectangular protuberance, the same SOV technique, together with the boundary conditions (\ref{5-020})-\ref{5-030}),  give rise to the field representations:
\begin{equation}\label{6-030}
u^{[1]}(\mathbf{x})=\sum_{m=0}^{\infty}\left[a_{m}\exp\big(ik_{ym}^{[1]}y\big)+b_{m}\exp\big(-ik_{ym}^{[1]}y\big)\right]
f_{m}(x) ~,
\end{equation}
\begin{equation}\label{6-040}
u^{[2]}(\mathbf{x})=\sum_{m=0}^{\infty}d_{m}\cos\big[k_{ym}^{[2]}(y-h)\big]g_{m}(x)~,
\end{equation}
wherein:
\begin{equation}\label{6-045}
f_{m}(x)=\exp(ik_{xm}x)~~,~~g_{m}(x)=\cos[k_{xm}(x+w/2)]~,
\end{equation}
\begin{equation}\label{6-050}
k_{xm}=\frac{m\pi}{w}~~,~~k_{ym}^{[l]}=\sqrt{\big(k^{[l]}\big)^{2}-\big(k_{xm}\big)^{2}}~~,~~\Re k_{ym}^{[l]}\ge 0~~,~~\Im k_{ym}^{[l]}\ge 0~;~\omega\ge 0 ~,~l=1,2~.
\end{equation}
The functions $f(k_{x},x)$ and $g_{m}(x)$ satisfy the orthogonality relations:
\begin{equation}\label{6-060}
\frac{1}{2\pi}\int_{-\infty}^{\infty}f(k_{x},x)f(-K_{x},x)dx=\delta(k_{x}-K_{x})~~;~~K_{x}\in\mathbb{R}~,                                                           ~,
\end{equation}
\begin{equation}\label{6-070}
\frac{1}{w}\int_{-w/2}^{w/2}g_{m}(x)g_{l}(x)dx=\frac{\delta_{lm}}{\epsilon_{l}}~~;~~l=0,1,2,...~,                                                           ~,
\end{equation}
in which $\delta(k_{x}-K_{x})$ is the Dirac delta distribution,
$\delta_{lm}$ the Kronecker delta symbol, and $\epsilon_{l}$ the Neumann symbol.
\subsection{Employment of the SOV field representations in the remaining boundary and continuity conditions}
The boundary condition (\ref{5-010}) and the continuity condition (\ref{5-050}) entail
\begin{equation}\label{7-010}
\frac{\mu^{[0]}}{2\pi}\int_{-\infty}^{\infty}u_{,y}^{[0]}(x,0)f(-K_{x},x)dx=
\frac{\mu^{[1]}}{2\pi}\int_{-w/2}^{w/2}u_{,y}^{[1]}(x,0)f(-K_{x},x)dx~~;~~\forall K_{x}\in\mathbb{R}~,
\end{equation}
whereas the  the continuity conditions (\ref{5-040}), (\ref{5-060}) and (\ref{5-070}) give rise to:
\begin{equation}\label{7-020}
\frac{1}{w}\int_{-w/2}^{w/2}u^{[0]}(x,0)g_{l}(x)dx=
\frac{1}{w}\int_{-w/2}^{w/2}u^{[1]}(x,0)g_{l}((x)dx~;~l=0,1,2,....~,
\end{equation}
\begin{equation}\label{7-030}
\frac{1}{w}\int_{-w/2}^{w/2}u^{[1]}(x,h_{1})g_{l}(x)dx=
\frac{1}{w}\int_{-w/2}^{w/2}u^{[2]}(x,h_{1})g_{l}((x)dx~;~l=0,1,2,....~,
\end{equation}
\begin{equation}\label{7-040}
\frac{\mu^{[1]}}{w}\int_{-w/2}^{w/2}u_{,y}^{[1]}(x,h_{1})g_{l}(x)dx=
\frac{\mu^{[2]}}{w}\int_{-w/2}^{w/2}u_{,y}^{[2]}(x,h_{1})g_{l}((x)dx~;~l=0,1,2,....~.
\end{equation}
The introduction of the field SOV representations into these four expressions lead, after use is made of the orthogonality relations (\ref{6-060})-(\ref{6-070}), to:
\begin{equation}\label{7-050}
a_{l}+b_{l}=2a^{i}\epsilon_{l}I_{l}^{+}(k_{x}^{i})+
\epsilon_{l}\int_{-\infty}^{\infty}\mathcal{B}(k_{x})I_{l}^{+}(k_{x})\frac{dk_{x}}{k^{[0]}}~;~l=0,1,2,....~,
\end{equation}
\begin{equation}\label{7-060}
\mathcal{B}(K_{x})=-\frac{w}{2\pi}\frac{\mu^{[1]}}{\mu^{[0]}}
\sum_{m=0}^{\infty}(a_{m}-b_{m})k_{ym}^{[1]}I_{m}^{-}(K_{x})~;~\forall K_{x}\in\mathbb{R}~,
\end{equation}
\begin{equation}\label{7-070}
a_{l}\exp(ik_{yl}^{[1]}h_{1})+b_{l}\exp(-ik_{yl}^{[1]}h_{1})=d_{l}\cos(k_{yl}^{[2}h_{2})
~;~l=0,1,2,....~,
\end{equation}
\begin{equation}\label{7-080}
a_{l}\exp(ik_{yl}^{[1]}h_{1})-b_{l}\exp(-ik_{yl}^{[1]}h_{1})=
d_{l}\frac{1}{i}\frac{\mu^{[2]}}{\mu^{[1]}}\frac{k_{yl}^{[2]}}{k_{l}^{[1]}}\sin(k_{yl}^{[2}h_{2})
~;~l=0,1,2,....~,
\end{equation}
in which $K_{y}^{[1]}=\sqrt{\big(k^{[1]}\big)^{2}-\big(K_{x}\big)^{2}}$ and:
\begin{equation}\label{7-090}
I_{m}^{\pm}(k_{x})=\int_{-w/2}^{w/2}\exp(\pm ik_{x}x)\cos[k_{xm}(x+w/2)]\frac{dx}{w}
~,
\end{equation}
and it is easy to show that
\begin{equation}\label{7-100}
I_{m}^{\pm}(k_{x})=\frac{i^{m}}{2}\text{sinc}\big[(\pm k_{x}+k_{xm})\frac{w}{2}\big]+\frac{(-i)^{m}}{2}\text{sinc}\big[(\pm k_{x}-k_{xm})\frac{w}{2}\big]
~.
\end{equation}

We thus have at our disposal four coupled expressions (i.e., (\ref{7-050})-(\ref{7-080}) which should enable us to determine the four sets of unknowns $\{\mathcal{B}(k_{x})\}$, $\{a_{m}\}$, $\{b_{m}\}$, $\{d_{m}\}$.
\subsection{Explicit expressions for each of the four sets of unknowns}\label{explicit}
Eqs. (\ref{7-070})-(\ref{7-080}) readily yield:
\begin{equation}\label{8-010}
a_{l}=d_{l}\left(\frac{\exp(-ik_{yl}^{[1]}h_{1})}{2i\mu^{[1]}k_{yl}^{[1]}}\right)
\left[i\mu^{[1]}k_{yl}^{[1]}\cos(k_{yl}^{[2}h_{2})+\mu^{[2]}k_{yl}^{[2]}\sin(k_{yl}^{[2}h_{2})\right]~;~l=0,1,2,....~,
\end{equation}
\begin{equation}\label{8-020}
b_{l}=d_{l}\left(\frac{\exp(ik_{yl}^{[1]}h_{1})}{2i\mu^{[1]}k_{yl}^{[1]}}\right)
\left[i\mu^{[1]}k_{yl}^{[1]}\cos(k_{yl}^{[2}h_{2})-\mu^{[2]}k_{yl}^{[2]}\sin(k_{yl}^{[2}h_{2})\right]~;~l=0,1,2,....~,
\end{equation}
whence
\begin{equation}\label{8-030}
a_{l}+b_{l}=d_{l}\kappa_{l}~~,~~a_{l}-b_{l}=-id_{l}\sigma_{l}~;~l=0,1,2,....~,
\end{equation}
with
\begin{equation}\label{8-040}
\kappa_{l}=\cos(k_{yl}^{[1]}h_{1})\cos(k_{yl}^{[2]}h_{2})-
\frac{\mu^{[2]}k_{yl}^{[2]}}{\mu^{[1]}k_{yl}^{[1]}}\sin(k_{yl}^{[1]}h_{1})\sin(k_{yl}^{[2]}h_{2})~,
\end{equation}
\begin{equation}\label{8-050}
\sigma_{l}=\sin(k_{yl}^{[1]}h_{1})\cos(k_{yl}^{[2]}h_{2})+
\frac{\mu^{[2]}k_{yl}^{[2]}}{\mu^{[1]}k_{yl}^{[1]}}\cos(k_{yl}^{[1]}h_{1})\sin(k_{yl}^{[2]}h_{2})~.
\end{equation}
Finally,the introduction of (\ref{8-030}) into (\ref{7-050})-\ref{7-060})results in:
\begin{equation}\label{7-050}
d_{l}\frac{\kappa_{l}}{\epsilon_{l}}=2a^{i}I_{l}^{+}(k_{x}^{i})+
\int_{-\infty}^{\infty}\mathcal{B}(k_{x})I_{l}^{+}(k_{x})\frac{dk_{x}}{k^{[0]}}~;~l=0,1,2,....~,
\end{equation}
\begin{equation}\label{8-070}
\mathcal{B}(k_{x})=\frac{iw}{2\pi}\frac{\mu^{[1]}}{\mu^{[0]}}
\sum_{m=0}^{\infty}d_{m}^{(M)}\sigma_{m}k_{ym}^{[1]}I_{m}^{-}(k_{x})~;~\forall k_{x}\in\mathbb{R}~.
\end{equation}
%
\subsection{Approximations of the sets of equations}
Until now everything has been rigorous provided the equations in the statement of the boundary-value problem are accepted as the true expression of what is involved in the seismic response of the protuberance.  In order to actually solve for the sets $\{d_{m}\}$, and then for  $\{a_{m}\}$, $\{b_{m}\}$, $\{d_{m}\}$, $\{\mathcal{B}(k_{x})\}$ (each of whose populations were  considered to be infinite until now) we must now resort  to approximations.

The approach is basically to replace (\ref{7-010})-(\ref{7-080}) by the finite system of linear equations
\begin{equation}\label{8-080}
a_{l}^{(M)}+b_{l}^{(M)}=2a^{i}\epsilon_{l}I_{l}^{+}(k_{x}^{i})+
\epsilon_{l}\int_{-\infty}^{\infty}\mathcal{B}^{(M)}(k_{x})I_{l}^{+}(k_{x})\frac{dk_{x}}{k^{[0]}}~;~l=0,1,2,...M.~,
\end{equation}
\begin{equation}\label{8-085}
\mathcal{B}^{(M)}(K_{x})=-\frac{w}{2\pi}\frac{\mu^{[1]}}{\mu^{[0]}}
\sum_{m=0}^{M}(a_{m}^{(M)}-b_{m}^{(M)})k_{ym}^{[1]}I_{m}^{-}(K_{x})~;~\forall K_{x}\in\mathbb{R}~,
\end{equation}
\begin{equation}\label{8-090}
a_{l}^{(M)}\exp(ik_{yl}^{[1]}h_{1})+b_{l}^{(M)}\exp(-ik_{yl}^{[1]}h_{1})=d_{l}^{(M)}\cos(k_{yl}^{[2}h_{2})
~;~l=0,1,2,....M~,
\end{equation}
\begin{equation}\label{8-095}
a_{l}^{(M)}\exp(ik_{yl}^{[1]}h_{1})-b_{l}^{(M)}\exp(-ik_{yl}^{[1]}h_{1})=
d_{l}^{(M)}\frac{1}{i}\frac{\mu^{[2]}}{\mu^{[1]}}\frac{k_{yl}^{[2]}}{k_{l}^{[1]}}\sin(k_{yl}^{[2}h_{2})
~;~l=0,1,2,....M~,
\end{equation}
from which we obtain, as by the previous steps,
\begin{equation}\label{8-100}
a_{l}^{(M)}+b_{l}^{(M)}=d_{l}^{(M)}\kappa_{l}~~,~~a_{l}^{(M)}-b_{l}^{(M)}=-id_{l}^{(M)}\sigma_{l}~;~l=0,1,2,....M~,
\end{equation}
and
\begin{equation}\label{8-110}
\mathcal{B}^{(M)}(k_{x})=\frac{iw}{2\pi}\frac{\mu^{[1]}}{\mu^{[0]}}
\sum_{m=0}^{M}d_{m}^{(M)}\sigma_{m}k_{ym}^{[1]}I_{m}^{-}(k_{x})~;~\forall k_{x}\in\mathbb{R}~.
\end{equation}
\begin{equation}\label{8-120}
d_{l}^{(M)}\frac{\kappa_{l}}{\epsilon_{l}}=2a^{i}I_{l}^{+}(k_{x}^{i})+
\int_{-\infty}^{\infty}\mathcal{B}^{(M)}(k_{x})I_{l}^{+}(k_{x})\frac{dk_{x}}{k^{[0]}}~;~l=0,1,2,....M~,
\end{equation}
in which the superscript  $(M)$ signifies the $M$-th order approximation of the indicated quantity, the  procedure being to increase $M$ so as to generate the sequence of  solutions for $M=0$, $M=1$, etc. until the values of the first few members of  these sets stabilize and the remaining members become very small.

It is important to underline the fact that the approximate solutions $\{a_{m}^{(M)}\}$, $\{b_{m}^{(M)}\}$, $\{d_{m}^{(M)}\}$, $\{\mathcal{B}^{(M)}(k_{x})\}$, together with the associated approximate field representations
\begin{equation}\label{8-120}
u^{s(M)}(\mathbf{x})=\int_{-\infty}^{\infty}\mathcal{B}^{(M)}(k_{x})f(k_{x},x)\exp(-ik_{y}y)\frac{dk_{x}}{k^{[0]}_{y}}~,
\end{equation}
\begin{equation}\label{8-130}
u^{[1](M)}(\mathbf{x})=\sum_{m=0}^{M}\left[a_{m}^{(M)}\exp\big(ik_{ym}^{[1]}y\big)+b_{m}^{(M)}\exp\big(-ik_{ym}^{[1]}y\big)\right]
f_{m}(x) ~,
\end{equation}
\begin{equation}\label{8-140}
u^{[2](M)}(\mathbf{x})=\sum_{m=0}^{M}d_{m}^{(M)}\cos\big[k_{ym}^{[2]}(y-h)\big]g_{m}(x)~,
\end{equation}
satisfy all the conditions of the boundary-value problem (i.e., Helmholtz equations, radiations condition, boundary conditions and continuity conditions) {\it for every $M\ge 0$}, but the associated field representations are mathematically not complete which is the reason why these solutions are qualified as approximate. We show hereafter that, in spite of this incomplete nature of our approximate solutions, the latter satisfy the conservation of flux relation for all $M\ge 0$.
\subsection{Final step in the demonstration of the conservation of flux}
On account of the last statement, the demonstration of the conservation of flux law for all $M\ge 0$ amounts to showing that the relation
\begin{equation}\label{9-010}
\mathcal{J}^{(M)}+\mathcal{K}^{(M)}=\mathcal{L}^{(M)}~;~\forall M=0,1,2,...~,
\end{equation}
is satisfied.

Let us begin with $\mathcal{L}^{(M)}$, which, due to the continuity relations across $\Gamma_{m}$, is
\begin{equation}\label{9-020}
\mathcal{L}^{(M)}=\frac{\mu^{[1]}}{2\mu^{[0]}}\Im\int_{-w/2}^{w/2}u^{[1](M)*}(x,0)u_{,y}^{[1](M)}(x,0)dx=-\frac{1}{2}\Im\int_{-w/2}^{w/2}u^{[0](M)*}(x,0)u_{,y}^{[0](M)}(x,0)dx~.
\end{equation}
The key point is to invoke the boundary conditions on $\Gamma_{l}$ and $\Gamma_{r}$ so as to obtain
\begin{equation}\label{9-030}
\mathcal{L}^{(M)}=--\frac{1}{2}\Im\int_{-\infty}^{\infty}u^{[0](M)*}(x,0)u_{,y}^{[0](M)}(x,0)dx~.
\end{equation}
Eq. (\ref{8-120}) informs us that:
\begin{equation}\label{9-040}
u^{[0](M)*}(x,0)=2u^{i*}(x,0)+u^{s(M)*}(x,0)=
\int_{-\infty}^{\infty}\Big[2a^{i*}\delta(k_{x}-k_{x}^{i})+\frac{\mathcal{B}^{(M)*}(k_{x})}{k^{[0]*}_{y}}\Big]\exp(-ik_{x}x)dk_{x}~,
\end{equation}
and
\begin{equation}\label{9-050}
u_{,y}^{[0](M)}(x,0)=u_{,y}^{s(M)}(x,0)=-i\int_{-\infty}^{\infty}\mathcal{B}^{(M)}(k_{x})\exp(ik_{x}x)dk_{x}~,
\end{equation}
whence, after the interchange of integrals,
\begin{equation}\label{9-060}
\mathcal{L}^{(M)}=
-\frac{1}{2}\Im\int_{-\infty}^{\infty}dk_{x}\Big[2a^{i*}\delta(k_{x}-k_{x}^{i})+\frac{\mathcal{B}^{(M)*}(k_{x})}{k_{y}^{[0]*}}\Big]
\int_{-\infty}^{\infty}dk'_{x}\frac{\mathcal{B}^{(M)}(k_{x})}{ik_{y}^{[0]}}\int_{-\infty}^{\infty}dx\exp[i(k_{x}-k'(x))x]~.
\end{equation}
However,
\begin{equation}\label{9-070}
\int_{-\infty}^{\infty}\exp[i(k_{x}-k'(x))x]=2\pi\delta(k_{x}-k'_{x})dx~.
\end{equation}
whence, by use of the sifting property of the Dirac delta distribution,
\begin{equation}\label{9-080}
\mathcal{L}^{(M)}=\Re\left[2\pi a^{i*}\mathcal{B}^{(M)}(k_{x}^{i})+\pi\int_{-\infty}^{\infty}\|\mathcal{B}^{(M)}(k_{x})\|^{2}\frac{dk_{x}}{k_{y}^{[0]*}}\right]~.
\end{equation}
With the change of variables $k_{x}=k^{[0]}\cos\phi$ and $k_{x}^{i}=k^{[0]}\cos\phi^{i}$, and making use of previously-evoked definitions  $\mathcal{B}(k^{[0]}\cos\phi)=\frac{B(\phi)}{\pi}$, and $\mathcal{B}(k^{[0]}\cos\phi^{i})=\frac{B(\phi^{i})}{\pi}=\frac{B\left(\theta^{i}+\frac{3\pi}{2}\right)}{\pi}$, we get
\begin{equation}\label{9-090}
\mathcal{L}^{(M)}=2\Re a^{i*}B^{(M)}\left(\theta^{i}+\frac{3\pi}{2}\right)+ \frac{1}{\pi}\int_{\pi}^{2\pi}\|B^{(M)}(\phi)\|^{2}d\phi~,
\end{equation}
and since, by definition, $2\Re a^{i*}B^{(M)}\left(\theta^{i}+\frac{3\pi}{2}\right)=\mathcal{K}^{(M)}$ and $\frac{1}{\pi}\int_{\pi}^{2\pi}\|B^{(M)}(\phi)\|^{2}d\phi=\mathcal{J}^{(M)}$, we conclude that
\begin{equation}\label{9-100}
\mathcal{L}^{(M)}=\mathcal{J}^{(M)}+\mathcal{K}^{(M)}~;~\forall M=0,1,2,...~,
\end{equation}
which means that our formulation, which involves only the continuity relations across $\Gamma_{m}$ and the plane wave representation of $u^{[0](M)}$, is such as to obey the conservation of flux law for all $M\ge0$.

A perhaps more-convincing demonstration of this result is obtained in the following manner. The starting point is
\begin{equation}\label{9-120}
\mathcal{L}^{(M)}=-\frac{\mu^{[1]}}{2\mu^{[0]}}\Im\int_{-w/2}^{w/2}u^{[1](M)*}(x,0)u_{,y}^{[1](M)}(x,0)dx~.
\end{equation}
Eq. (\ref{8-140}) informs us that
\begin{equation}\label{9-130}
u^{[1]*(M)}(x,0)=\sum_{m=0}^{M}\left[a_{m}^{(M)*}+b_{m}^{(M)*}\right]
f_{m}(x) ~,
\end{equation}
and
\begin{equation}\label{9-140}
u_{,y}^{[1](M)}(x,0)=\sum_{m=0}^{M}ik_{ym}^{[1]}\left[a_{m}^{(M)*}-b_{m}^{(M)*}\right]
f_{m}(x) ~,
\end{equation}
which, on account of (\ref{8-100}), give rise to
\begin{equation}\label{9-150}
u^{[1]*(M)}(x,0)=\sum_{m=0}^{M}d_{m}^{(M)*}\kappa_{m}^{*}f_{m}(x)~,
\end{equation}
and
\begin{equation}\label{9-160}
u_{,y}^{[1](M)}(x,0)=\sum_{m=0}^{M}d_{m}^{(M)}k_{ym}^{[1]}\sigma_{m}f_{m}(x)~.
\end{equation}
It follows, after sum and integral changes, that
\begin{equation}\label{9-170}
\mathcal{L}^{(M)}=-\frac{\mu^{[1]}}{2\mu^{[0]}}\Im\sum_{m=0}^{M}d_{m}^{(M)*}\kappa_{m}^{*}
\sum_{n=0}^{M}d_{n}^{(M)}k_{yn}^{[1]}\sigma_{n}\int_{-w/2}^{w/2}dx f_{m}(x)f_{n}~,
\end{equation}
or, on account of the orthogonality relation relative to $f_{m}$,
\begin{equation}\label{9-180}
\mathcal{L}^{(M)}=-\frac{\mu^{[1]}}{2\mu^{[0]}}\Im\sum_{m=0}^{M}d_{m}^{(M)*}\kappa_{m}^{*}
\sum_{n=0}^{M}d_{n}^{(M)}k_{yn}^{[1]}\sigma_{n}w\frac{\delta_{mn}}{\epsilon_{m}}~,
\end{equation}
which, because of the sifting property of the Kronecker delta, becomes
\begin{equation}\label{9-190}
\mathcal{L}^{(M)}=-w\frac{\mu^{[1]}}{2\mu^{[0]}}\Im\sum_{m=0}^{M}\|d_{m}^{(M)*}\|^{2}\frac{\kappa_{m}^{*}
k_{ym}^{[1]}\sigma_{m}}{\epsilon_{m}}~.
\end{equation}
Now, by virtue of the continuity of displacement across $\Gamma_{m}$ and the boundary condition on $\Gamma_{l}$ and $\Gamma_{r}$, we get
\begin{equation}\label{9-200}
\int_{-\infty}^{\infty}\left[2u^{i}(x,0)+u^{s(M)}(x,0)\right]\exp(-ik_{x}x)dx=\int_{-w/2}^{w/2}u^{[1](M)}(x,0)\exp(-ik_{x}x)dx~;~\forall k_{x}\in\mathbb{R}~,
\end{equation}
which, by virtue of (\ref{9-040}), (\ref{9-130}), (\ref{9-150}), (\ref{9-070}), and the sifting property of the Dirac delta distribution, takes the form
\begin{equation}\label{9-210}
4\pi a^{i}\delta(k_{x}^{i}-k_{x})+\frac{2\pi}{k_{y}^{[0]}}\mathcal{B}^{(M)}(k_{x})=
w\sum_{m=0}^{M}d_{m}^{(M)}\kappa_{m}I_{m}^{-}(k_{x})~;~\forall k_{x}\in\mathbb{R}~,
\end{equation}
from which it follows that
\begin{multline}\label{9-220}
\int_{-\infty}^{\infty}\|\mathcal{B}^{(M)}(k_{x})\|^{2}\frac{dk_{x}}{k_{y}^{[0]*}}=\\
-2a^{i*}\int_{-\infty}^{\infty}\mathcal{B}^{(M)}(k_{x})\delta(k_{x}^{i}-k_{x})dk_{x}+
\frac{iw}{2\pi}\int_{-\infty}^{\infty}dk_{x}\mathcal{B}^{(M)}(k_{x})\sum_{m=0}^{M}\frac{w}{2\pi}d_{m}^{(M)*}\kappa_{m}^{*}I_{m}^{+}(k_{x})
~,
\end{multline}
or on account of (\ref{8-110})
\begin{multline}\label{9-230}
\int_{-\infty}^{\infty}\|\mathcal{B}^{(M)}(k_{x})\|^{2}\frac{dk_{x}}{k_{y}^{[0]*}}=\\
-2a^{i*}\mathcal{B}^{(M)}(k_{x}^{i})+
\frac{iw}{(2\pi)^{2}}\frac{\mu^{[1]}}{\mu^{[0]}}\sum_{m=0}^{M}d_{m}^{(M)*}\kappa_{m}^{*}\sum_{n=0}^{M}d_{n}^{(M)*}\sigma_{n}k_{n}^{[1]}
\int_{-\infty}^{\infty}dk_{x}I_{m}^{+}(k_{x})I_{n}^{-}(k_{x})
~.
\end{multline}
Owing to the orthogonality relations satisfied by $f(k_{x},x)$ and $f_{m}(x)$ we find
\begin{equation}\label{9-240}
\int_{-\infty}^{\infty}I_{m}^{+}(k_{x})I_{n}^{-}(k_{x})dk_{x}=\frac{2\pi}{w}\frac{\delta_{mn}}{\epsilon_{m}}~,
\end{equation}
so that
\begin{equation}\label{9-250}
\int_{-\infty}^{\infty}\|\mathcal{B}^{(M)}(k_{x})\|^{2}\frac{dk_{x}}{k_{y}^{[0]*}}=
-2a^{i*}\mathcal{B}^{(M)}(k_{x}^{i})+
\frac{iw}{2\pi}\frac{\mu^{[1]}}{\mu^{[0]}}\sum_{m=0}^{M}\|d_{m}^{(M)}\|^{2}\frac{\kappa_{m}^{*}\sigma_{n}k_{n}^{[1]}}{\epsilon_{m}}
~.
\end{equation}
Taking the real part of this expression finally yields
\begin{equation}\label{9-260}
\pi\int_{-k^{[0]}}^{k^{[0]}}\|\mathcal{B}^{(M)}(k_{x})\|^{2}\frac{dk_{x}}{k_{y}^{[0]}}+
\pi\Re\left[2a^{i*}\mathcal{B}^{(M)}(k_{x}^{i})\right]=
-\frac{w}{2}\frac{\mu^{[1]}}{\mu^{[0]}}\Im\sum_{m=0}^{M}\|d_{m}^{(M)}\|^{2}\frac{\kappa_{m}^{*}\sigma_{n}k_{n}^{[1]}}{\epsilon_{m}}
~.
\end{equation}
which, on account of (\ref{9-190}) and the change of variables $k_{x}=k^{[0]}\cos\phi$ again yields
\begin{equation}\label{9-270}
\frac{1}{\pi}\int_{\pi}^{2\pi}\|B^{(M)}(\phi)\|^{2}d\phi+2\Re\left[a^{i*}B^{(M)}(\theta^{i}+\frac{3\pi}{2})\right]=\mathcal{L}^{(M)}
~,
\end{equation}
wherein we recognize that  $2\Re \left[a^{i*}B^{(M)}(\theta^{i}+\frac{3\pi}{2})\right]=\mathcal{K}^{(M)}$ and $\frac{1}{\pi}\int_{\pi}^{2\pi}\|B^{(M)}(\phi)\|^{2}d\phi=\mathcal{J}^{(M)}$, so that once again we obtain

\begin{equation}\label{9-280}
\mathcal{J}^{(M)}+\mathcal{K}^{(M)}=\mathcal{L}^{(M)}~;~\forall M=0,1,2,...
~,
\end{equation}
thus showing that the $M$-th order approximate solutions indeed satisfy the conservation of flux law  for all $M\ge 0$.

It is important to stress that this conclusion: (i) is valid for all $M\ge 0$, and (ii) is valid both in the absence of losses (in which case $\mathcal{L}^{(M)}=0$) and the presence of losses (in which case $\mathcal{L}^{(M)}\ne 0$).
\section{Verifications of the conservation of flux law by means of the numerical solutions for the case of a lossy or lossless protuberance of rectangular shape}
Until now, we treated the the forward-scattering problem in a formal manner, i.e., based on the formal (i.e., not numerical) solution of the four equations (\ref{8-080})-(\ref{8-095}). We now address the problem of how to actually (i.e., numerically) solve this system of four equations and then employ the numerical solutions to see if they are such as to (numerically) satisfy the conservation of flux relation.
\subsection{Ingredients of the numerical method}
We already showed that this system can be reduced to  two coupled systems of equations (\ref{8-110})-(\ref{8-120}), which we re-write here for convenience:
\begin{equation}\label{10-010}
\mathcal{B}^{(M)}(k_{x})=\frac{iw}{2\pi}\frac{\mu^{[1]}}{\mu^{[0]}}
\sum_{m=0}^{M}d_{m}^{(M)}\sigma_{m}k_{ym}^{[1]}I_{m}^{-}(k_{x})~;~\forall k_{x}\in\mathbb{R}~.
\end{equation}
\begin{equation}\label{10-020}
d_{l}^{(M)}\frac{\kappa_{l}}{\epsilon_{l}}=2a^{i}I_{l}^{+}(k_{x}^{i})+
\int_{-\infty}^{\infty}\mathcal{B}^{(M)}(k_{x})I_{l}^{+}(k_{x})\frac{dk_{x}}{k^{[0]}}~;~l=0,1,2,....M~,
\end{equation}

The final trick is to introduce(\ref{10-010}) into (\ref{10-020}) so as to give rise to the system of linear equations for the $\{d_{l}\}$
\begin{equation}\label{10-030}
\sum_{m=0}^{M}E_{lm}^{(M)}d_{m}^{(M)}=c_{l}~;~l=0,1,2,...M~,
\end{equation}
in which:
\begin{equation}\label{10-040}
E_{lm}^{(M)}=\delta_{lm}\frac{\kappa_{l}}{\epsilon_{l}}-
\frac{iw}{2\pi}\frac{\mu^{[1]}}{\mu^{[0]}}k_{ym}^{[1]}\sigma_{m}J_{lm}~~,~~c_{l}=2a^{i}I_{l}^{+}(k_{x}^{i})
~~,~~J_{lm}=\int_{-\infty}^{\infty}I_{l}^{+}(k_{x})I_{m}^{-}(k_{x})\frac{dk_{x}}{k_{y}^{[0]}}~.
\end{equation}

As it stands,  the matrix equation $\mathbf{E}^{(M)}\mathbf{d}^{(M)}=\mathbf{c}$ is not particularly-appropriate for the determination of the diffraction coefficient vector $\mathbf{d}^{(M)}$. The reason for this is that certain elements of the matrix $\mathbf{E}^{(M)}$ become very large for large $M$ so as to make the intversion of $\mathbf{E}^{(M>>1)}$ problematic.

The way to resolve this problem is actually quite simple: in (\ref{10-030}), divide $E_{lm}$ by $\sigma_{m}$ and multiply $d_{m}$ by $\sigma_{m}$ so as to obtain
\begin{equation}\label{10-050}
\sum_{m=0}^{\infty}\mathcal{E}_{lm}^{(M)}\mathcal{F}_{m}^{(M)}=\mathcal{G}_{l}~;~l=0,1,2,...~,
\end{equation}
in which:
\begin{equation}\label{10-060}
\mathcal{E}_{lm}=E_{lm}\frac{\epsilon_{l}}{\sigma_{m}}=\delta_{lm}\frac{\kappa_{l}}{\sigma_{l}}-
\epsilon_{l}\frac{iw}{2\pi}\frac{\mu^{[1]}}{\mu^{[0]}}k_{ym}^{[1]}J_{lm}~~,~~
\mathcal{G}_{l}=c_{l}\epsilon_{l}=2a^{i}\epsilon_{l}I_{l}^{+}(k_{x}^{i})
~~,~~\mathcal{F}_{m}=d_{m}^{(M)}\sigma_{m}~.
\end{equation}
From the numerical point of view, there is now no difficulty in solving for $\mathcal{F}_{m}$ via (\ref{10-050}).

The  fulll numerical procedure is then to increase $M$ so as to generate the sequence of numerical solutions $\{F_{0}^{(0)}\}$, $\{F_{0}^{(1)},F_{1}^{(1)}\}$,....until the values of the first few members of  these sets stabilize and the remaining members become very small. This is usually obtained for reasonably-small values of $M$, especially in the low frequency regime of interest in our seismic response problem.

 The problem that was ignored until now is that of providing a suitable means  for numerically evaluating $J_{lm}$. This problem  be treated either in the manners explained in (\cite{wi73,wi90}) or as follows. From the definition (\ref{7-090}) of $I_{lm}^{\pm}$ and that (\ref{3-100}) of the Hankel function, it follows, after changes in the order of integration, that
\begin{equation}\label{10-070}
J_{lm}=\frac{\pi}{w^{2}}\int_{-w/2}^{w/2}dx'\cos[k_{xl}(x'+w/2)]\int_{-w/2}^{w/2}dx\cos[k_{xm}(x+w/2)]H_{0}^{(1)}(k^{[0]}|x'-x|)
~,
\end{equation}
this double integral (over finite limits) being accessible to standard (e.g., Simpson) quadrature techniques.
\subsection{Tests of the conservation of flux relation pertaining to the numerical solutions}
All the following numerical examples apply to the case of a  small (in height) double-layer hill submitted to an obliquely-incident seismic plane body wave: $a^{i}=1~ (a.u.)$, $\theta^{i}=70^{\circ}$, $h_{1}=75~m$, $h_{2}=75~m$, $w=750~m$, $\mu^{[0]}=6.85~GPa$, $\beta^{[0]}=1629.4~ms^{-1}$,  $\beta^{[1]}=1300-i10~ms^{-1}$, $\mu^{[2]}=2~GPa$, $\beta^{[2]}=1000-i10~ms^{-1}$. Thus, the only changes from one example to another concern $M$, $\mu^{[1]}$ or $f$.
\clearpage
\newpage
\subsubsection{Effect of the variation of $M$ for  fixed frequency   and shear modulus  in $\Omega_{1}$}
\begin{figure}[ht]
\begin{center}
\includegraphics[width=0.65\textwidth]{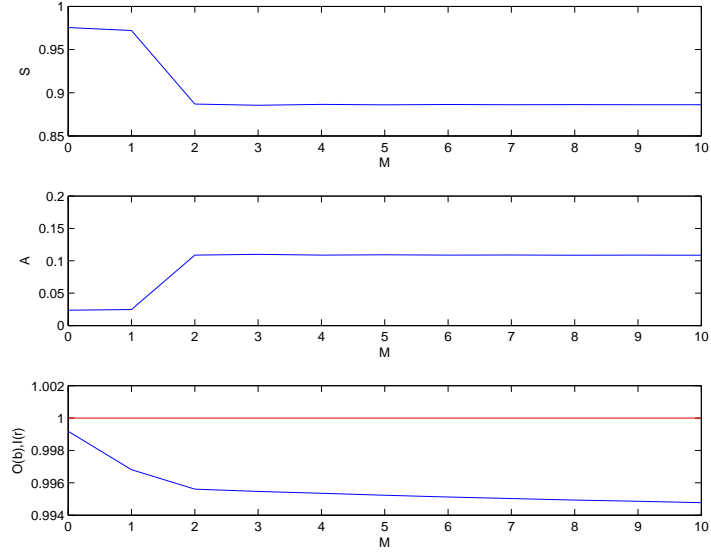}
\caption{The three panels depict fluxes as a function of  $M$. The upper, middle and lower panels are relative to the normalized scattered flux $\mathcal{S}^{(M)}$, the normalized absorbed flux $\mathcal{A}^{(M)}$ and the normalized output flux $\mathcal{O}^{(M)}=\mathcal{S}^{(M)}+\mathcal{A}^{(M)}$ respectively. The blue curves are all the result of numerical solutions and the red line in the lower panel depicts the normalized input flux $\mathcal{I}=1$ which is the goal for $O$ if the conservation law $\mathcal{O}=\mathcal{I}$ is to be satisfied. Case $f=2~Hz$ and $\mu^{[1]}=4~GPa$}
\label{hill-010}
\end{center}
\end{figure}

Fig. \ref{hill-010} is relative to a variation of $M$.
\clearpage
\newpage
Although the values of $\mathcal{S}^{(M)}$ and $\mathcal{A}^{(M)}$ are seen (at least graphically) to stabilize starting with $M=2$, ${O}^{(M)}$ is seen to depart progressively (although slightly) from $\mathcal{I}=1$, the reason for this being the numerical errors in the computation of $J_{lm}$ (because when the latter quantities are computed with more accuracy it was found that the difference of ${O}^{(M)}$ from ${I}=1$ decreases). In spite of this, the conservation law is seen to be satisfied with an error of less than a half percent.

 An important feature of fig. \ref{hill-010} is that the conservation law is numerically satisfied for all $M$ even though the lower-order (i.e., $M<2$) solutions are in obvious error (this meaning that they are not stabilized, assuming that the correct values of $\mathcal{S}^{[M]}$ and $\mathcal{A}^{[M]}$ are their stabilized (large $M$) values).
\subsection{Effect of the variation of the frequency $f$ for various $M$ and $\mu^{[1]}$}
\begin{figure}[ht]
\begin{center}
\includegraphics[width=0.65\textwidth]{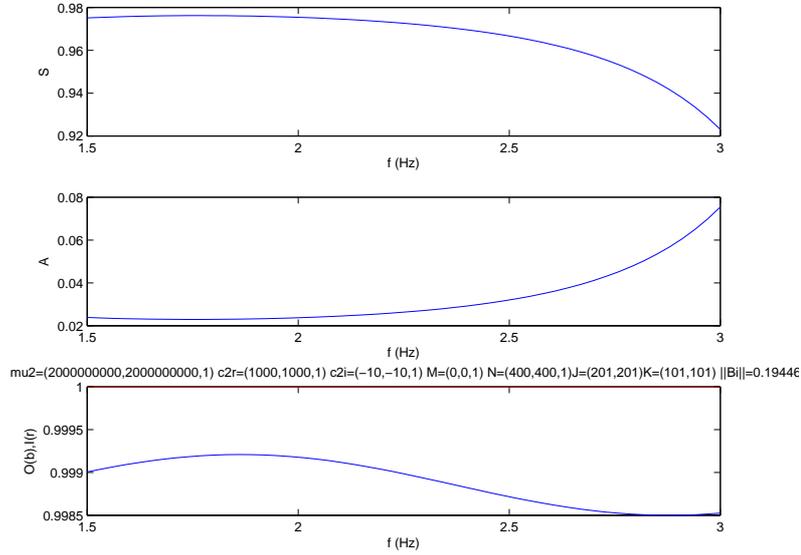}
\caption{The three panels depict fluxes as a function of  $f$. The upper, middle and lower panels are relative to the normalized scattered flux $\mathcal{S}^{(M)}$, the normalized absorbed flux $\mathcal{A}^{(M)}$, and the normalized output flux $\mathcal{O}^{(M)}=\mathcal{S}^{(M)}+\mathcal{A}^{(M)}$ respectively. The blue curves are all the result of numerical solutions and the red line in the lower panel depicts the normalized input flux $\mathcal{I}=1$ which is the goal for $O$ if the conservation law $\mathcal{O}=\mathcal{I}$ is to be satisfied. Case $M=0$, $\mu^{[1]}=4~GPa$.}
\label{hill-020}
\end{center}
\end{figure}

\begin{figure}[ptb]
\begin{center}
\includegraphics[width=0.65\textwidth]{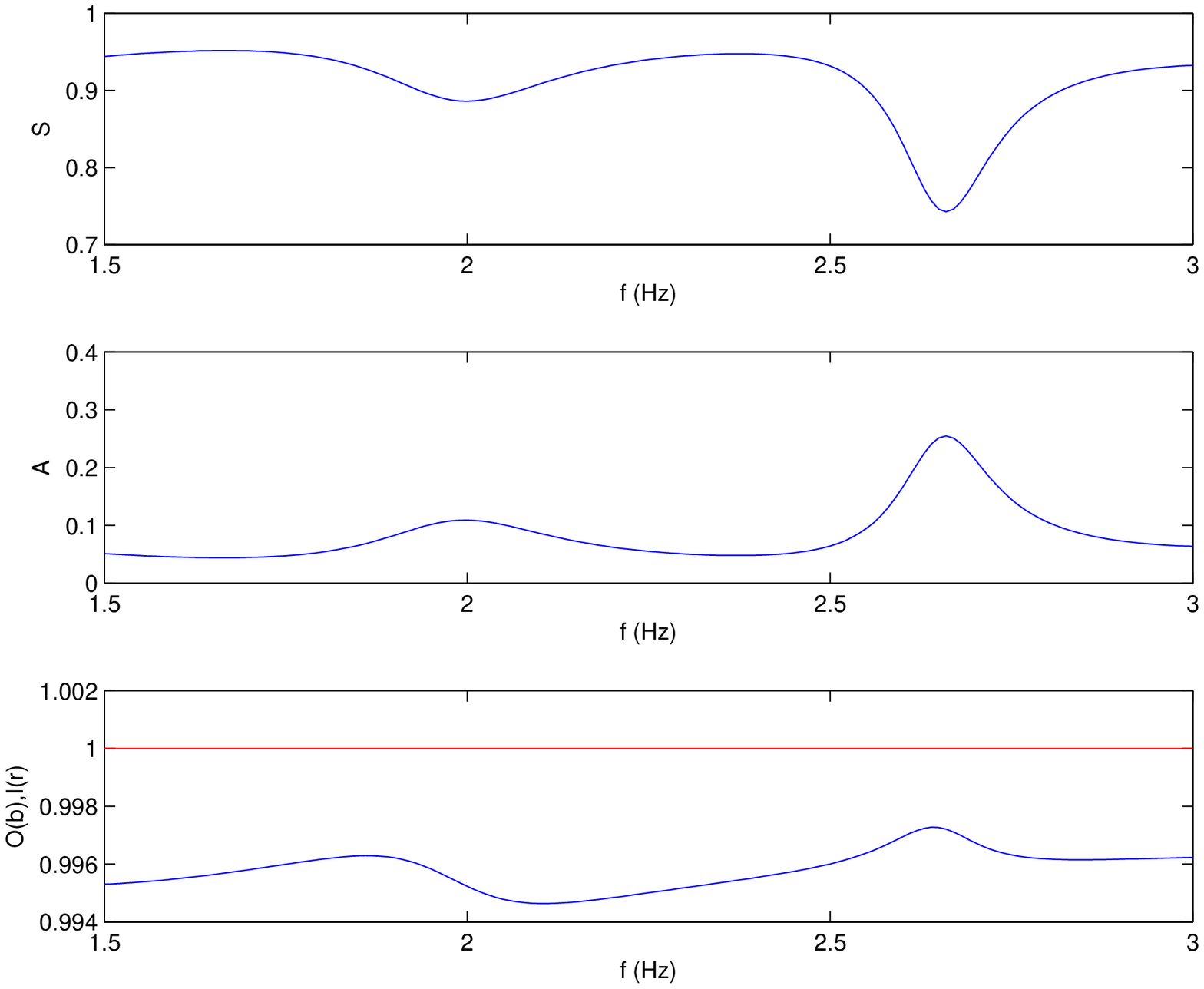}
\caption{Same as fig. \ref{hill-020} except that now $M=5$  and $\mu^{[1]}=4~GPa$.}
\label{hill-030}
\end{center}
\end{figure}

\begin{figure}[ptb]
\begin{center}
\includegraphics[width=0.65\textwidth]{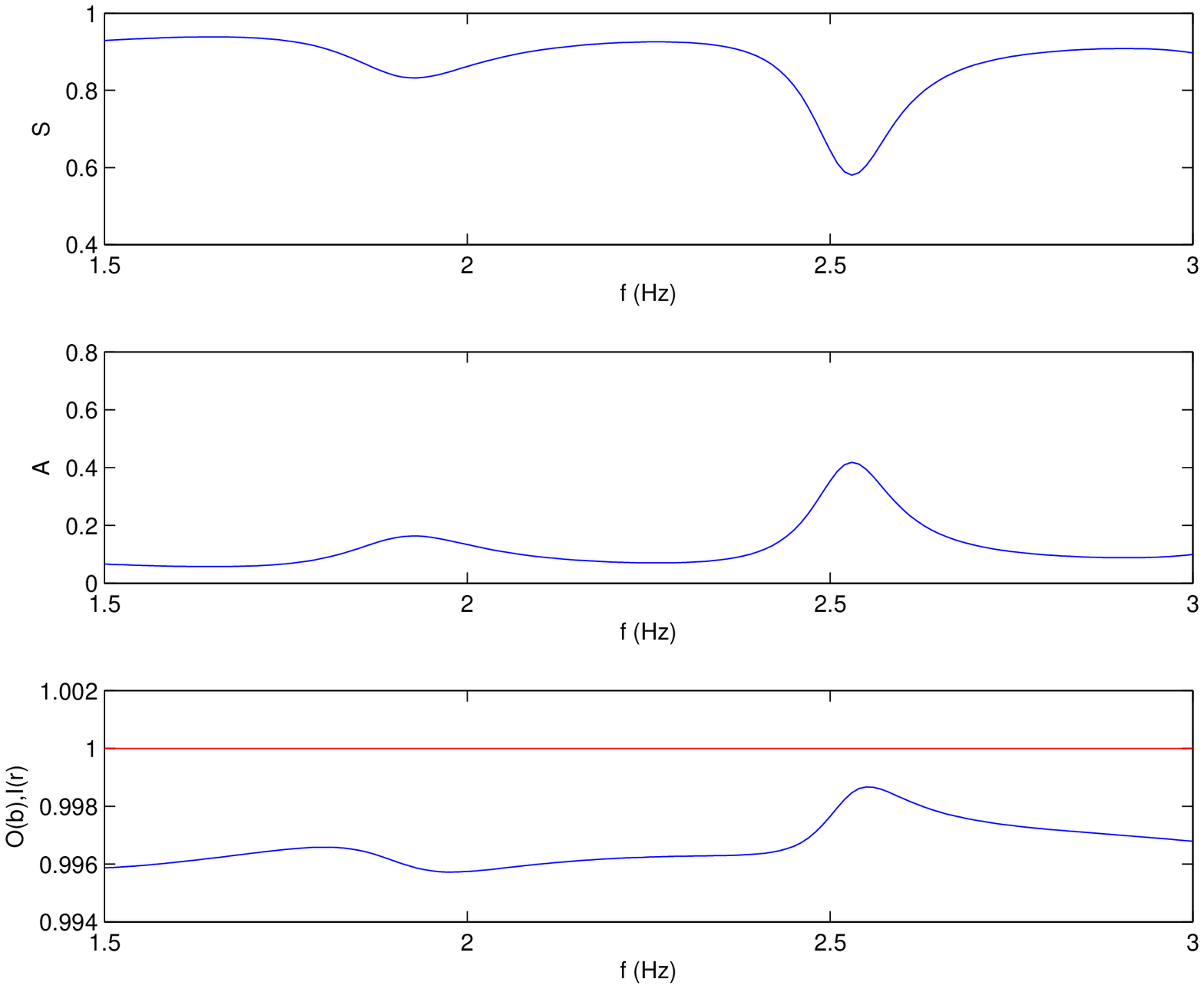}
\caption{Same as fig. \ref{hill-020} except that now $M=5$ and $\mu^{[1]}=2~GPa$.}
\label{hill-040}
\end{center}
\end{figure}

Figs. \ref{hill-020}-\ref{hill-040} are  relative to a variations of $f$.
\clearpage
\newpage
In these three examples,  the conservation law is again seen to be satisfied with an error of less than a half percent for all $M$. The comparison of  fig. \ref{hill-020} (relative to the $M=0$ solution) and  fig. \ref{hill-030} (relative to the $M=5$ stabilized, presumably correct, solution) shows that the $M=0$ solution is acceptable only at very low frequencies $f<1.7~Hz$. This latter finding is bad news for those who offer approximate solutions similar to our $M=0$ solution as explanations of how an above-ground structure responds, over a rather large range of frequencies, to a seismic wave.
\section{Conclusions}
This study was motivated by the necessity of disposing of a tool for testing the validity of theoretical and numerical models of a class of scattering problems of considerable interest in seismology:  the response of manmade (e.g., building) or natural (e.g., hill) topographical feature(s) emerging from flat ground to a seismic wave.  We demonstrated mathematically (i.e., which makes use solely of the equations inherent in the boundary-value problem together with Green's theorem) the existence of a general conservation law for the problem of the scattering of seismic waves by a protuberance occupied by a lossy or non-lossy bilayer. We showed that this law, which applies to a protuberance of arbitrary shape, takes the form of the simple relation $\mathcal{O}=\mathcal{I}$, or $\mathcal{S}+\mathcal{A}=\mathcal{I}$ wherein $\mathcal{O}$ is the normalized output flux, $\mathcal{I}=1$ the normalized input flux, $\mathcal{S}$  the normalized scattered flux (often termed radiation damping) and $\mathcal{A}$ the normalized absorbed flux (which vanishes when the medium within the protuberance is non-lossy).  Our contention was that for the above-mentioned model(s) be valid, it(they) should at least be such as to satisfy the conservation of flux law.

To show the usefulness of this law, we applied it to the specific case of a rectangular cylinder bilayer  protuberance submitted to a plane body SH wave. This problem lends itself to a domain decomposition separation of variables (DD-SOV) analysis which enables the formal solution for the amplitudes of the waves in the various subdomains of the configuration to be exhibited in the form of a system of  coupled matrix equations, each matrix being of infinite dimensions. By making solely use of this system of equations we showed that the latter is such as to verify exactly the conservation of flux relation. The limitation of the order of the matrices to a finite value $M$ forms the basis of a method of obtaining approximate solutions; the latter were also shown to satisfy exactly the conservation of flux law. This, of course, raised the question: if an approximate solution (which is trivially not exact) exactly satisfies the conservation of flux law, is  this law really useful for deciding non-ambiguously whether a given solution to the scattering problem is valid?

To give some insight to this question, we outlined a method for obtaining explicit (the previous system of equations yielding only implicit or formal solutions) numerical solutions to the rectangular protuberance scattering problem. The scheme for solving the problem was first based on the demonstration that the aformentioned system of  matrix equations can be reduced to a single infinite-order matrix equation, the unknowns of which are the amplitudes of the wavefield in the uppermost layer of the rectangular protuberance.  A sequence of $M$-th order (for $M=0,1,2,...$) now-numerical  solutions were shown to be easily-extracted from the single linear system and these solutions were submitted to the conservation of flux test. Naturally, it was expected that the lower-order solutions be further-removed from the true solution than the higher-order ones, and it was hoped that these solutions stabilize for large-enough $M$, under the hypothesis that the stabilized solution is the closest to the true solution of the scattering problem. We found that the $M$-th order numerical solution satisfied the conservation of flux relation to within a half percent for all $M$, even though the low-order solutions were manifestly far from the true (in the aforementioned sense) counterparts, a finding consistent with what was previously found for the formal solutions.

All this leads to the conclusion that the satisfaction of a conservation of flux law (like that of any conservation (e.g., energy) law) is a necessary, but not sufficient, condition for a solution of the scattering problem to be qualified as true.

\end{document}